REVIEW ARTICLE

# Prospects for observing and localizing gravitational-wave transients with Advanced LIGO, Advanced Virgo and KAGRA

**Abbott, B. P. et al. (KAGRA Collaboration, LIGO Scientific Collaboration and Virgo Collaboration)***

**Abstract**

We present our current best estimate of the plausible observing scenarios for the Advanced LIGO, Advanced Virgo and KAGRA gravitational-wave detectors over the next several years, with the intention of providing information to facilitate planning for multi-messenger astronomy with gravitational waves. We estimate the sensitivity of the network to transient gravitational-wave signals for the third (O3), fourth (O4) and fifth observing (O5) runs, including the planned upgrades of the Advanced LIGO and Advanced Virgo detectors. We study the capability of the network to determine the sky location of the source for gravitational-wave signals from the inspiral of binary systems of compact objects, that is binary neutron star, neutron star–black hole, and binary black hole systems. The ability to localize the sources is given as a sky-area probability, luminosity distance, and comoving volume. The median sky localization area (90% credible region) is expected to be a few hundreds of square degrees for all types of binary systems during O3 with the Advanced LIGO and Virgo (HLV) network. The median sky localization area will improve to a few tens of square degrees during O4 with the Advanced LIGO, Virgo, and KAGRA (HLVK) network. During O3, the median localization volume (90% credible region) is expected to be on the order of $10^5, 10^6, 10^7$ Mpc$^3$ for binary neutron star, neutron star–black hole, and binary black hole systems, respectively.

This article is a revised version of https://doi.org/10.1007/s41114-018-0012-9.

**Change summary** Major revision, updated and expanded.

**Change details** Since publication of the previous version (Abbott et al 2018f), several updates to the document have been made. The most significant changes are that we now frame our projections in terms of observing runs, we include final results from O2, and we updated our localization projections to include KAGRA as a fourth detector. Key differences are outlined in the Appendix.

✉ KAGRA Collaboration, LIGO Scientific Collaboration and Virgo Collaboration
kscboard-chair@icrr.u-tokyo.ac.jp; lsc-spokesperson@ligo.org; virgo-spokesperson@ego-gw.it

*The full author list and affiliations are given at the end of paper.

The localization volume in O4 is expected to be about a factor two smaller than in O3. We predict a detection count of $1^{+12}_{-1}(10^{+52}_{-10})$ for binary neutron star mergers, of $0^{+19}_{-0}(1^{+91}_{-1})$ for neutron star–black hole mergers, and $17^{+22}_{-11}(79^{+89}_{-44})$ for binary black hole mergers in a one-calendar-year observing run of the HLV network during O3 (HLVK network during O4). We evaluate sensitivity and localization expectations for unmodeled signal searches, including the search for intermediate mass black hole binary mergers.

**Keywords** Gravitational waves · Gravitational-wave detectors · Electromagnetic counterparts · Data analysis

# Contents

## 1 Introduction

Advanced LIGO (Aasi et al. 2015a), Advanced Virgo (Acernese et al. 2015), and KAGRA (Somiya 2012; Aso et al. 2013) are kilometer-scale gravitational-wave (GW) detectors that are sensitive to GWs with frequencies of $\sim$20–2000 Hz.[1] The era of GW astronomy began with the detection of GW150914 (Abbott et al. 2016i), a signal from the coalescence of a binary black hole (BBH); the first confirmed multi-messenger counterpart to a GW observation came with GW170817 (Abbott et al. 2017i), a signal from a binary neutron star (BNS) coalescence which was accompanied by detections across the electromagnetic spectrum (Abbott et al. 2017j). In this article, we describe the schedule, sensitivity, sky-localization accuracy, and expected detections for the GW-detector network. We discuss the past, present, and future planned sequence of observing runs and the prospects for multi-messenger astronomy.

The purpose of this article is to provide information to the astronomy community to assist in the formulation of plans in the era of GW observations. In particular, we intend this article to provide the information required for assessing the features of programs for joint observation of GW events using electromagnetic, neutrino, or other facilities (e.g., Abbott et al. 2016h, 2017j; Adrian-Martinez et al. 2016; Albert et al. 2017a, b).

The full science of ground-based GW detectors is broad (Abbott et al. 2018e), and is not covered in this article. We concentrate solely on candidate GW transient signals. We place particular emphasis on the coalescence of binary systems of compact objects, such as BNS and neutron star–black hole (NSBH) systems, which are the GW sources for which electromagnetic follow-up is most promising (Goodman 1986; Paczynski 1986; Eichler et al. 1989; Li and Paczynski 1998; Kulkarni 2005; Rosswog 2005; Metzger et al. 2010; Roberts et al. 2011; Abadie et al. 2012b, c; Evans et al. 2012; Metzger and Berger 2012; Nissanke et al. 2013; Kasen et al. 2013; Barnes and Kasen 2013; Tanaka and Hotokezaka 2013; Aasi et al. 2014a; Grossman et al. 2014; Ciolfi and Siegel 2015; Ghirlanda et al. 2016; Paschalidis 2017; Rosswog et al. 2017; Foucart et al. 2018; Barbieri et al. 2019; Metzger 2020), and BBHs, which are the most commonly detected source (Abbott et al. 2016b, 2017f, 2018a, c). No electromagnetic emission is expected for vacuum BBH mergers (Centrella et al. 2010), but is possible if there is surrounding material (Schnittman 2013), for example, remnants of mass lost from the parent star (Perna et al. 2016; Janiuk et al. 2017) or if the binary was embedded in a common envelope (Woosley 2016), or a disk of an active galactic nucleus (Bartos et al. 2017; Stone et al. 2017). Mergers of binary systems of compact objects are absolute distance indicators, and thus can be used as standard sirens to estimate the Hubble constant (Schutz 1986; Holz and Hughes 2005; Dalal et al. 2006; Nissanke et al. 2010; Abbott et al. 2017a). When an electromagnetic counterpart, and hence a host galaxy cannot be identified, a statistical approach which uses galaxy catalogs and the GW localization volume can be used (Del Pozzo 2012; Chen et al. 2018; Fishbach et al. 2019; Soares-Santos et al. 2019). For more general introductory articles on GW generation, detection and astrophysics, we point readers to Blanchet (2014), Pitkin et al. (2011) and Sathyaprakash and Schutz (2009).

[1] LIGO is short for Laser Interferometer Gravitational-Wave Observatory. KAGRA is named after the Japanese word *KAGURA*, which means traditional sacred music and dance for the gods; the name has a secondary meaning as an abbreviation for KAmioka GRavitational-wave Antenna. Virgo is named for the Virgo constellation and is not written in capital letters.

As the detector network grows and evolves we will release updated versions of this article: This is the fourth version. The plausible observing scenarios for the upcoming observing runs includes KAGRA and the upgrades of the Advanced LIGO (aLIGO) and Advanced Virgo (AdV) detectors, called A+ and AdV+, respectively. The predicted sky-localization accuracies and detection rates have been updated and now incorporate the atsrophysical results from the first and second observing runs (Abbott et al. 2018a, c). Changes with respect to the previous version (Aasi et al. 2016) are listed in Appendix A. Throughout the paper we assume a flat cosmology with Hubble parameter $H_0 = 67.9$ km $s^{-1}Mpc^{-1}$, and density parameters $\Omega_m = 0.3065$ and $\Omega_\Lambda = 0.6935$ (Ade et al. 2016).

## 2 Construction, commissioning and observing phases

We divide the development of the GW observatories into three phases:

**Construction:** includes the installation and testing of the detectors. This phase ends with *acceptance* of the detectors. Acceptance means that the interferometers can lock for periods of hours: light is resonant in the arms of the interferometer with *no guaranteed GW sensitivity*. Construction incorporates several short *engineering runs* with no astrophysical output as the detectors progress towards acceptance. The aLIGO construction project ended in March 2015. The construction of AdV was completed in early 2017. Construction of KAGRA will be completed by mid-late 2019.

**Commissioning:** improves the detectors' performance with the goal of reaching design sensitivity. Engineering runs in the commissioning phase allow us to understand our detectors and analyses in an observational mode; these are not intended to produce astrophysical results, but that does not preclude the possibility of this happening.[2] Rather than proceeding directly to design sensitivity before making astrophysical observations, commissioning is interweaved with *observing runs*.

**Observing:** begins when the detectors have reached (and can stably maintain) a significantly improved sensitivity compared with previous operation. Observing runs produce astrophysical results such as direct detections from certain GW sources and upper limits on the rates or energetics of others. During the first two observing runs (O1 and O2) a Memorandum Of Understanding (MOU) governed the exchange of GW candidates between astronomical partners and the LIGO and Virgo Collaborations. From the start of the third observing run (O3) GW event candidates identified in low-latency are released immediately to the full astronomical community (see Sect. 4 for details). KAGRA will become a part of the global network with full data sharing in the latter half of O3.

[2] The detection of GW150914 occurred in the engineering run ER8 immediately preceding the formal start of O1.

Commissioning is a complex process which involves both scheduled improvements to the detectors and tackling unexpected new problems. While our experience makes us cautiously optimistic regarding the schedule for the advanced detectors, it is not possible to make concrete predictions for sensitivity or duty cycle as a function of time.

As a standard figure of merit for detector sensitivity, we use the range, $R$, evaluated for CBCs consisting of representative masses. We define $V$ as the orientation-averaged spacetime volume surveyed per unit detector time, assuming a matched-filter detection signal-to-noise ratio (SNR) threshold of 8 in a single detector. The volume $V$ corresponds to the comoving volume with the inclusion of a $(1+z)$ factor to account for time dilation (redshifted volume $V_z$ in Chen et al. 2017). For a population of sources with a constant comoving source-frame rate density, $V$ multiplied by the rate density gives the detection rate of those sources by the particular detector. The range $R$ is obtained as $(4\pi/3)R^3 = V$. For further insight into the range, and a discussion of additional quantities such as the median and average distances to sources, see (Chen et al. 2017).

For unmodeled short-duration ($\lesssim 1$ s) signals or bursts, we evaluate an approximate sensitive luminosity distance determined by the total energy $E_{\mathrm{GW}}$ emitted in GWs, the central frequency $f_0$ of the burst, the detector noise power spectral density $S(f_0)$, and the single-detector SNR threshold $\rho_{\mathrm{det}}$ (Sutton 2013):

$$D \simeq \left( \frac{G}{2\pi^2 c^3} \frac{E_{\mathrm{GW}}}{S(f_0) f_0^2 \rho_{\mathrm{det}}^2} \right)^{1/2}. \quad (1)$$

This distance is then corrected by the time dilation cosmology factor to obtain the surveyed volume $V$, and the range $R$.

### 2.1 O1: aLIGO

O1 began on 18 September 2015 and ended on 12 January 2016. Data from the surrounding engineering periods were of sufficient quality to be included in the analysis, meaning that observational data was collected from 12 September 2015 to 19 January 2016. The run involved the Hanford (H) and Livingston (L) detectors (Abbott et al. 2016e; Martynov et al. 2016). We aimed for a BNS range of 60–80 Mpc for both instruments (see Fig. 1), and achieved a 80 Mpc range.

The localizations of the three BBH events detected during this run (GW150914, GW151012,[3] GW151226), exhibit the characteristic broken arc for a two-detector network (Abbott et al. 2016b, h, 2018c). GW150914 and GW151226 were shared with partner astronomers soon after detection. Their poor localization (the 90% credible regions are given in Table 3) made the follow-up challenging (Abbott et al.

[3] The significance of LVT151012, initially classified as a GW candidate, increased after reanalysis of the O1 data with improved detection pipelines. It is now considered an astrophysical GW event (Abbott et al. 2018c).

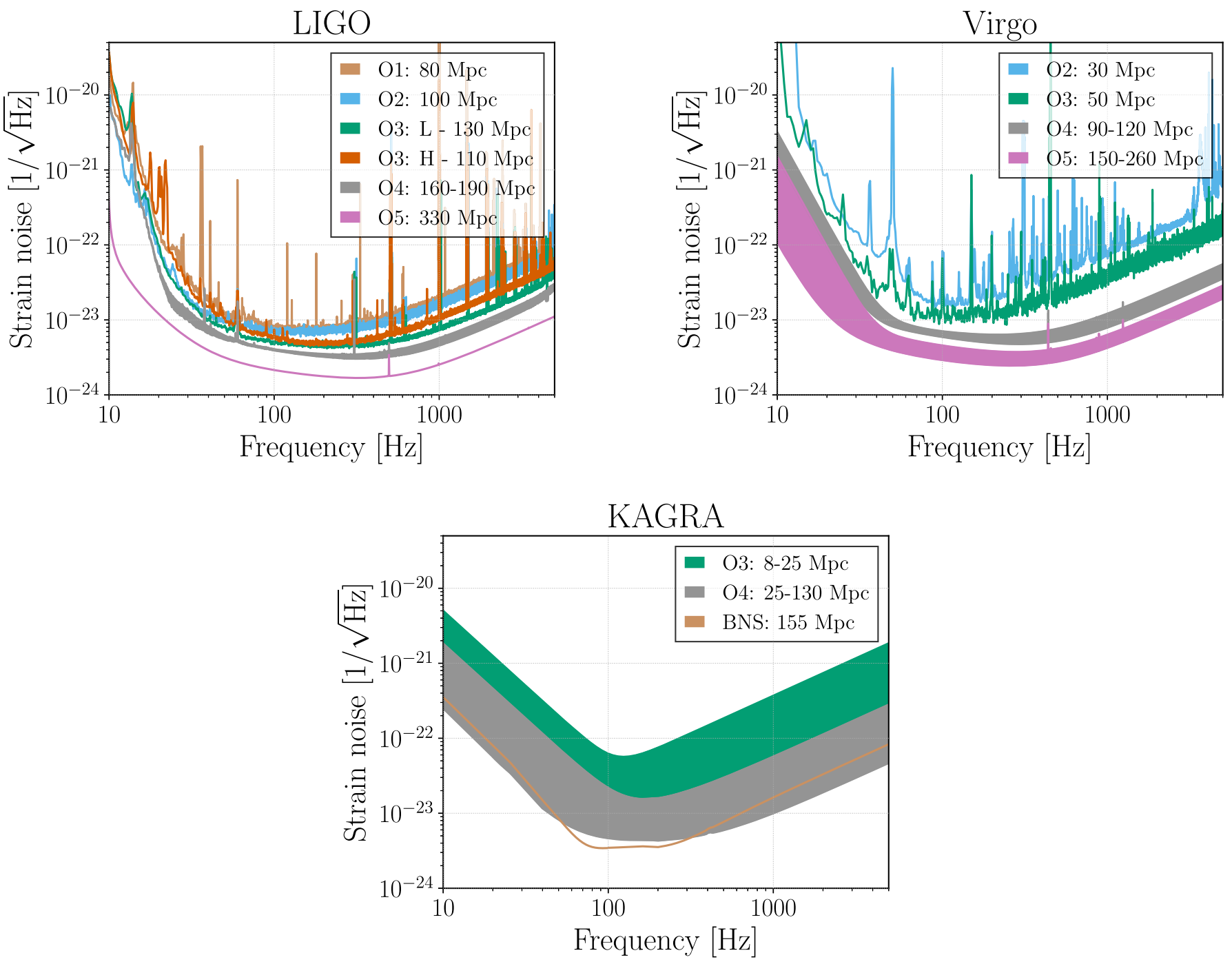

**Fig. 1** aLIGO (top left), AdV (top right) and KAGRA (bottom) target strain sensitivities as a function of frequency. The quoted range is for a $1.4\,M_\odot + 1.4\,M_\odot$ BNS merger. The BNS range (in megaparsec) achieved in past observing runs and anticipated for future runs is shown. The O1 aLIGO curve is taken from the Hanford detector, the O2 aLIGO curve comes from Livingston. In each case these had the better performance for that observing run. The O3 curves for aLIGO and AdV reflect recent performance. For some runs the anticipated ranges are shown as bands reflecting the uncertainty in the impact of improvements and upgrades to the overall sensitivity. Detailed planning for the post-O3 to O4 period is now in progress and may result in changes to both target sensitivities for O4 and the start date for this run. The KAGRA BNS curve may be realized by detuning the signal recycling cavity to significantly improve the BNS range to 155 Mpc once design sensitivity is reached

2016h, l; Adrian-Martinez et al. 2016; Albert et al. 2017b). See Sect. 3.3 for more discussion of the O1 and O2 follow-up program.

In O1 the largest non-observing periods for each detector were due to Locking and Environmental issues (see Table 1). Locking refers to the amount of time spent in bringing the interferometers from an uncontrolled state to their lowest noise configuration (Staley et al. 2014). Environmental effects include earthquakes, wind and the microseism noise arising from ocean storms (Effler et al. 2015; Abbott et al. 2016c). The latter two effects have seasonal variation, with the prevalence of storms being higher during the winter months. The Livingston detector has a greater sensitivity to microseism noise and to earthquakes than Hanford, mainly due to the local geophysical environment (Daw et al. 2004).

**Table 1** Percentage of time during the first and second observing runs that the aLIGO and AdV detectors spent in different operating modes as recorded by the on-duty operator

| | O1 | | O2 | | |
|---|---|---|---|---|---|
| | Hanford | Livingston | Hanford | Livingston | Virgo |
| *Operating mode %* | | | | | |
| Observing | 64.6 | 57.4 | 65.3 | 61.8 | 85.1 |
| Locking | 17.9 | 16.1 | 8.0 | 11.7 | 3.1 |
| Environmental | 9.7 | 19.8 | 5.8 | 10.1 | 5.6 |
| Maintenance | 4.4 | 4.9 | 5.4 | 6.0 | 3.1 |
| Commissioning | 2.9 | 1.6 | 3.4 | 4.7 | 1.1 |
| Planned engineering | 0.1 | 0.0 | 11.9 | 5.5 | – |
| Other | 0.4 | 0.2 | 0.2 | 0.2 | 2.0 |

Since several factors may influence detector operation at any given time, there is a certain subjectivity to the assignments. Maintenance includes a planned 4-h weekly period ($\sim$ 2.4% of the total), and unplanned corrective maintenance to deal with equipment or hardware failures. Coincident operation of the aLIGO detectors occurred $\sim$ 43% of the time in O1 and $\sim$ 46% in O2. After joining O2 on August 1 2017 AdV operated with a duty factor of approximately 85% until the end of the run on August 25 2017

## 2.2 O2: aLIGO joined by AdV

O2 began on 30 November 2016 and ended on 25 August 2017. It was preceded by an engineering run which began on 31 October 2016 at Livingston and on 14 November 2016 at Hanford. The delay at Hanford was to facilitate extra commissioning activities. The achieved sensitivity across the run was typically a BNS range of 80–100 Mpc (Abbott et al. 2018c).

The AdV interferometer (V; Acernese et al. 2015) joined O2 on 1 August 2017, forming a three detector network for the last month of the run. The goal was a BNS range of 40 Mpc. Because of a vacuum contamination issue, which has since been resolved, AdV used steel wires, rather than fused silica fibers, to suspend the test masses. This limited the highest possible BNS range for AdV; in O2 the BNS range achieved was 30 Mpc. The aLIGO and AdV sensitivities are shown in Fig. 1.

Of the eight GW signals detected during O2, five were localized by the three detector LIGO-Hanford, LIGO-Livingston and Virgo (HLV) network. From Table 3 we see that GW170818 was localized to a 90% credible region of 39 $\text{deg}^2$ making it the best localized BBH detection to date (Abbott et al. 2018c). GW170817, the first detection of a BNS merger, was localized to a 90% credible region of 16 $\text{deg}^2$. The enhanced accuracy is due to the addition of AdV to the network. The discoveries associated with this detection are highlighted in Sect. 3.3. An overview of the extensive multi-messenger observations accompanying GW170817 is given in Abbott et al. (2017j).

In O2 the aLIGO detectors saw some improvement in duty factors from operating during non-winter months, with an almost 50% reduction in the fraction of time lost to environmental effects at both sites (see Table 1). O2 also saw a rise in the fraction of time spent in planned engineering: it was a longer run and hence included a dedicated break in observations to effect needed repairs and to attempt improvements to the sensitivity. During O1 and O2, Livingston lost over twice as much observing time to earthquakes, microseism noise and wind compared to Hanford. For the aLIGO instruments improvements to control systems, the locking process, and the addition of extra sensors (Coughlin et al. 2017; Biscans et al. 2018; Ross et al. 2017; Venkateswara et al. 2014) may lead to modest increases in the duty factor of the aLIGO instruments. The Virgo instrument operated with a duty factor of approximately 85% after joining O2 and similar performance is expected during O3.

Our expectations from earlier versions of this document that we expect duty factors of at most 70–75% for each LIGO instrument during extended runs are borne out by experience. Assuming unplanned downtime periods are uncorrelated among detectors, these duty factor estimates imply that all detectors in a three-detector network will be operating in coincidence approximately 34– 42% of the time, and at least two detectors will be operating for 78– 84% of the time. For a four-detector network, three or more detectors will be operational around 65– 74% of the time, and for a five-detector network, three of more detectors will be operating for 84–90% of the time. The weekly maintenance period for aLIGO instruments overlaps for three of the 4 h. The timezone difference makes overlapping the AdV and aLIGO maintenance periods impractical. Longer planned engineering interruptions may take place at the same time across the network, so these coincidence times are conservative estimates.

### 2.3 O3: aLIGO, AdV and KAGRA

The third observing run started on April 1, 2019 and was expected to end on April 30, 2020, with a commissioning break from October 1, 2019 to November 1, 2019. While this article was in review the COVID-19 Pandemic led to suspension of the observing run on March 27, 2020. The increase in sensitivity of the LIGO detectors (whose target sensitivity was expected to be 120 Mpc) comes from a variety of changes, chiefly from increasing the input laser power, adding a squeezed vacuum source at the interferometer output and mitigating noise arising from scattered light. Additionally, end test-mass optics with lower-loss coatings, along with new reaction masses, have been installed in each interferometer. The Livingston instrument began the run with an average BNS range of 130 Mpc and the Hanford instrument typically operates with an average range of 110 Mpc.

Fused silica fibers were installed on the AdV test mass suspensions in preparation for O3. Other improvements included reduction of technical noises, increasing the input laser power and installation of a squeezed vacuum source. The result was a BNS range of 50 Mpc at the start of O3.

The KAGRA detector (K; Somiya 2012; Aso et al. 2013) is located at the Kamioka underground site. The first operation of a detector in an initial configuration with a simple Michelson interferometer occurred in March 2016 (Akutsu et al. 2018). The detector is now being upgraded to its baseline design configuration. Initial operation was made in April–May 2018, in a simple Michelson configuration with a single end test mass cryogenically cooled to 20 K and the other

**Table 2** Achieved and projected detector sensitivities for a $1.4\,M_\odot + 1.4\,M_\odot$ BNS system, a $30\,M_\odot + 30\,M_\odot$ BBH system, a $1.4\,M_\odot + 10\,M_\odot$ NSBH system, and for two unmodeled burst signals

| | | O1 | O2 | O3 | O4 | O5 |
|---|---|---|---|---|---|---|
| BNS range (Mpc) | aLIGO | 80 | 100 | 110–130 | 160–190 | 330 |
| | AdV | – | 30 | 50 | 90–120 | 150–260 |
| | KAGRA | – | – | 8–25 | 25–130 | 130+ |
| BBH range (Mpc) | aLIGO | 740 | 910 | 990–1200 | 1400–1600 | 2500 |
| | AdV | – | 270 | 500 | 860–1100 | 1300–2100 |
| | KAGRA | – | – | 80–260 | 260–1200 | 1200+ |
| NSBH range (Mpc) | aLIGO | 140 | 180 | 190–240 | 300–330 | 590 |
| | AdV | – | 50 | 90 | 170–220 | 270–480 |
| | KAGRA | – | – | 15–45 | 45–290 | 290+ |
| Burst range (Mpc) [$E_{\mathrm{GW}} = 10^{-2}\,M_\odot c^2$] | aLIGO | 50 | 60 | 80–90 | 110–120 | 210 |
| | AdV | – | 25 | 35 | 65–80 | 100–155 |
| | KAGRA | – | – | 5–25 | 25–95 | 95+ |
| Burst range (kpc) [$E_{\mathrm{GW}} = 10^{-9}\,M_\odot c^2$] | aLIGO | 15 | 20 | 25–30 | 35–40 | 70 |
| | AdV | – | 10 | 10 | 20–25 | 35–50 |
| | KAGRA | – | – | 0–10 | 10–30 | 30+ |

The quoted ranges correspond to the orientation-averaged spacetime volumes surveyed per unit detector time. For the burst ranges, we assume an emitted energy in GWs at 140 Hz of $E_{\mathrm{GW}} = 10^{-2}\,M_\odot c^2$ and of $E_{\mathrm{GW}} = 10^{-9}\,M_\odot c^2$. The later is consistent with the order of magnitude of the energy expected from core-collapse of massive stars (see footnote 4). Both compact binary coalescence (CBC) and burst ranges are obtained using a single-detector SNR threshold of 8. The O1 and O2 numbers are representative of the best ranges for the LIGO detectors: Hanford in O1 and Livingston in O2. The O3 numbers for aLIGO and AdV reflect recent average performance of each of the three detectors. Range intervals are quoted for future observing runs due to uncertainty about the sequence and impact of upgrades test mass at room temperature. Subsequently, all the optical components have been installed and the test masses will be cryogenically cooled to reduce thermal noise. Early observations may come in late-2019–early 2020 with a range of 8–25 Mpc; KAGRA intends to join the network for the latter part of O3. The exact timing of observations has yet to be decided.

## 2.4 Commissioning and observing roadmap

The anticipated strain sensitivity evolution for aLIGO, AdV and KAGRA is shown in Fig. 1. In Table 2 we present values of the range for different detector networks and GW sources (BNSs, BBHs, NSBHs, and unmodelled signals, such as from the core-collpase of massive stars[4]). In previous versions of this paper, an option to optimize the detector sensitivity for a specific class of astrophysical signals, such as BNS mergers was discussed. Given the success of the aLIGO and AdV instruments and the approval of the new upgrades Advanced LIGO Plus (A+) and Advanced Virgo Plus (AdV+), such an optimization is no longer planned for these instruments.

[4] For details on different models of core-collapse supernovae, GW peak frequency and emitted energy, and the corresponding search sensitivities see e.g (Abbott et al. 2019c). Based on the majority of the theoretical expectations, it is unlikely that advanced detectors will be sufficiently sensitive to detect an extra-galactic core-collapse supernova.

Assuming that no unexpected obstacles are encountered, the aLIGO detectors are expected to achieve design sensitivity with a BNS range of 160–190 Mpc in O4. A configuration upgrade after O3 will increase the range of AdV to 90–120 Mpc in O4. KAGRA is currently intending participate fully in O4 with a BNS range of 25–130 Mpc. Owing to the cryogenic test mass suspension system, mirror coating thermal noise is expected to be lower than quantum noise. KAGRA will retain the option of optimizing the quantum noise by detuning the signal recycling cavity and significantly improve the BNS range to 155 Mpc.

Upgrading the existing instruments will enable LIGO and Virgo to increase their range with respect to the aLIGO and AdV detector design sensitivities. The A+ upgrade to the aLIGO instruments will include higher power, frequency-dependent squeezing and, crucially, new test masses with improved coating thermal noise. Facilities modifications to incorporate the filter cavity required for frequency-dependent squeezing will begin after O3. The full A+ configuration, adding improved test masses and balanced homodyne readout, is expected to be in place for O5. The AdV+ upgrade will occur in two phases. Phase 1 installation will begin after O3 and will involve adding signal recycling, frequency-dependent squeezing, higher input laser power (to 50 W from 20 W currently) and cancellation of Newtonian noise. Phase 2 will be implemented between O4 and O5 and will include input laser power increase to 200 W, 100 kg test masses and better optical coatings. Discussion of upgrades to increase the sensitivity of KAGRA in advance of O5 have begun, but the detailed plan and expected sensitivity are still being formulated.

The original aLIGO design called for three identical 4-km interferometers, two at Hanford and one at Livingston. In 2011, the LIGO Lab and the IndIGO[5] consortium in India proposed installing one of the aLIGO Hanford detectors at a new observatory in India (LIGO-India; Iyer et al. 2011). In early 2015, the LIGO Laboratory placed this interferometer in long-term storage for use in India. The Government of India granted in-principle approval to LIGO-India in February 2016. This detector will be configured, including upgrades, identically to the other LIGO instruments. Operation is anticipated in 2025.

GEO 600 (Lück et al. 2010; Dooley et al. 2016) will continue to operate as a GW detector beyond O3 as techniques for improving the sensitivity at high frequency are investigated (Affeldt et al. 2014). At its current sensitivity, it is unlikely to contribute to detections. By around 2021 with a deliberate focus on high frequency narrow-band sensitivity at a few kilohertz, GEO 600 may contribute to the understanding of BNS merger physics, as well as sky localization for such systems. In the meantime, it will continue observing with frequent commissioning and instrument science investigations related to detuned signal recycling and novel applications of squeezed light, as well as increasing the circulating power and levels of applied squeezing (Abadie et al. 2011a; Grote et al. 2013; Aasi et al. 2013a; Brown et al. 2017).

[5] www.gw-indigo.org.

Third-generation observatories, such as the Einstein Telescope[6] (Punturo et al. 2010), or Cosmic Explorer[7] (Abbott et al. 2017d), are envisioned in the future. It is also possible that for some sources, there could be multiband GW observations. The space-borne Laser Interferometer Space Antenna (LISA)[8] (Amaro-Seoane et al. 2017) could provide early warning and sky localization (Sesana 2016), as well as additional information on system parameters (Vitale 2016), formation mechanisms (Nishizawa et al. 2016a, b; Breivik et al. 2016) and tests of general relativity (Barausse et al. 2016). These future observatories are beyond the scope of this paper.

## 2.5 Envisioned observing schedule

Keeping in mind the important caveats about commissioning affecting the scheduling and length of observing runs, the following are plausible scenarios for the operation of the ground-based GW detector network over the next decade:

**2019–2020 (O3):** A year-long run (started April 1, 2019) with the aLIGO detectors at 110–130 Mpc and AdV at 50 Mpc. KAGRA plans to join for the latter part of the run with a range of 8–25 Mpc. A 1-month commissioning break for the LIGO and Virgo instruments is scheduled to begin October 1, 2019. To preserve the 12 month O3 observing period, the end date for O3 is now planned to be April 30, 2020. Possible extensions of the run will be limited so that O3 will end no later than June 30, 2020.

**Late 2021/Early 2022–Late 2022/Early 2023 (O4):** A four-detector network with the two aLIGO instruments at 160–190 Mpc; Phase 1 of AdV+ at 90–120 Mpc and KAGRA at 25–130 Mpc. The projected sensitivities and precise dates of this run are now being actively planned and remain fluid.

**Late 2024/Early 2025–2026 (O5):** O5 will begin with a four-detector network incorporating the A+ upgrade for the aLIGO instruments and the AdV+ Phase 2 upgrade for Virgo. The target range for aLIGO is 330 Mpc and for AdV it is 150–260 Mpc. KAGRA will operate at or above its O4 sensitivity of 130 Mpc.

**2025+:** With the addition of an upgraded aLIGO interferometer in India we will have a five-detector network: three aLIGO detectors with a design sensitivity of 330 Mpc, AdV at 150–260 Mpc and KAGRA at 130+ Mpc.

This timeline is summarized in Fig. 2.[9] Detailed planning for the post-O3 period is in progress and may result in significant changes to both target sensitivities and uncertainty in the start and end times of the planned observing runs, especially for those further in the future. As the network grows to include more detectors, sky localization will improve (Klimenko et al. 2011; Veitch et al. 2012; Nissanke et al. 2013; Rodriguez et al. 2014; Pankow et al. 2018), as will the fraction of observational time with multiple instruments on-sky. The observational implications of these scenarios are discussed in Sect. 5.

[6] www.et-gw.eu.

[7] www.cosmicexplorer.org.

[8] www.lisamission.org.

[9] GEO 600 will continue observing with frequent commissioning breaks during this period.

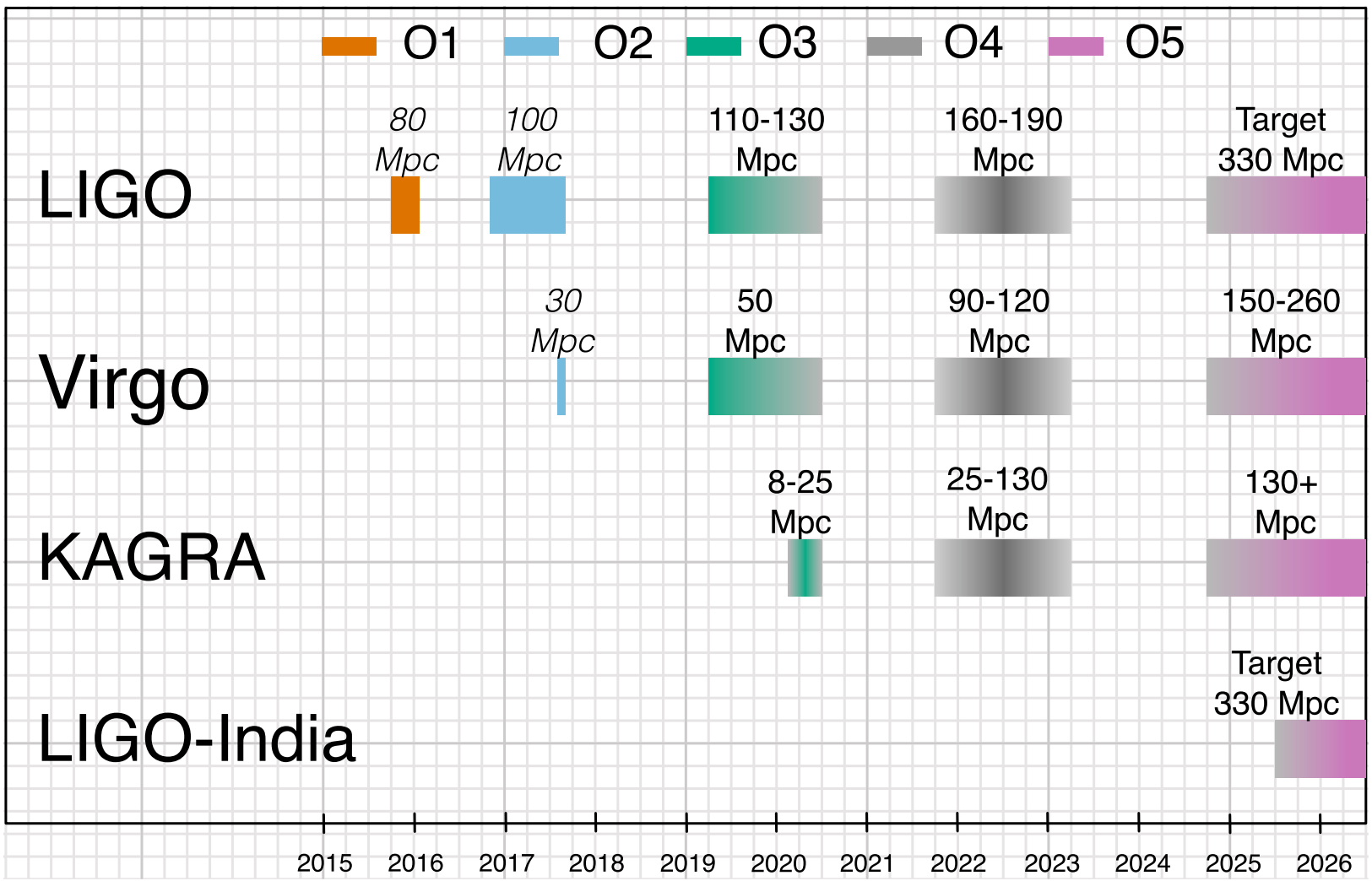

**Fig. 2** The planned sensitivity evolution and observing runs of the aLIGO, AdV and KAGRA detectors over the coming years. The colored bars show the observing runs, with achieved sensitivities in O1, O2 and O3, and the expected sensitivities given by the data in Fig. 1 for future runs. There is significant uncertainty in the start and end times of the planned observing runs, especially for those further in the future, and these could move forward or backwards relative to what is shown above. Uncertainty in start or finish dates is represented by shading. The break between O3 and O4 will last at least 18 months. O3 is expected to finish by June 30, 2020 at the latest. The O4 run is planned to last for one calendar year. We indicate a range of potential sensitivities for aLIGO during O4 depending on which upgrades and improvements are made after O3. The most significant driver of the aLIGO range in O4 is from the implementation of frequency-dependent squeezing. The observing plan is summarised in Sect. 2.5

## 3 Searches and localization of gravitational-wave transients

Data from GW detectors are searched for many types of possible signals (Abbott et al. 2018e). Here we focus on signals from CBCs, including BNS, NSBH and BBH systems and generic unmodeled transient signals.

Observational results of searches for transient signals are reported in detail elsewhere (Abbott et al. 2016b, d, j, o, 2017b, f, g, h, i, k, 2018c, g). The O1 and O2 results include ten clear detections originating from BBH coalescences and GW170817 which is the first detection of a BNS coalescence (Abbott et al. 2017i, 2018c). The public release of the LIGO and Virgo data allows researchers to perform independent analyses of the GW data. Some of these analyses report a few additional significant BBH event candidates (Zackay et al. 2019; Venumadhav et al. 2019, 2020). No other type of transient source has been identified during O1 and O2 (Abbott et al. 2016o, 2017b, l, 2018c).

Using the observation of GW170817, we estimate a BNS event rate of 110–3840 $\mathrm{Gpc}^{-3}$ $\mathrm{year}^{-1}$ (Abbott et al. 2018c). This rate is obtained by combining the results over different search pipelines and two different astrophysical populations, which assume a uniform mass distribution in the $1\,M_\odot$–$2\,M_\odot$ range for the NSs, and a Gaussian mass distribution (Özel and Freire 2016) centered at $1.33\,M_\odot$ with a standard deviation of $0.09\,M_\odot$. Compatible estimates for the merger rate were derived from the rate of electromagnetic transients similar to the counterpart of GW170817 (Siebert et al. 2017; Kasliwal et al. 2017; Smartt et al. 2017; Yang et al. 2017; Zhang et al. 2018). Rate estimation based upon astrophysical population models and observations of Galactic BNS systems remains an active area of research. The BNS merger rate inferred from O1 and O2 is close to the most optimistic values predicted by current astrophysical population models (e.g., Abadie et al. 2010b; Kim et al. 2013; Dominik et al. 2015; Vangioni et al. 2016; de Mink and Belczynski 2015; Eldridge et al. 2017, 2019; Belczynski et al. 2017; Kruckow et al. 2018; Mapelli and Giacobbo 2018; Giacobbo and Mapelli 2018; Barrett et al. 2018; Klencki et al. 2018; Spera et al. 2019; Pol et al. 2019; Chruslinska et al. 2019; Artale et al. 2019).

From the observations of BBHs during O1 and O2, we infer that their rate of mergers is 9.7–101 $\mathrm{Gpc}^{-3}$ $\mathrm{year}^{-1}$ (Abbott et al. 2018c). This rate combines results from different search pipelines and two astrophysical populations; a population of BBHs with primary mass following a power law distribution of index $\alpha = -2.3$, and a population of BBHs with primary mass distribution uniform in the log. For both populations, masses are cut off at a lower mass of $5\,M_\odot$ and at a maximum mass of $50\,M_\odot$ (Abbott et al. 2018a, c). Using a power law mass distribution with flexible values for the power law index, and the minimum and maximum masses (Model B in Abbott et al. 2018a), the BBH rate is estimated to be 25–109 $\mathrm{Gpc}^{-3}$ $\mathrm{year}^{-1}$. The non-detection of NSBHs in O1 and O2 allows us to place a 90% upper limit of the merger rate of 610 $\mathrm{Gpc}^{-3}$ $\mathrm{year}^{-1}$ (Abbott et al. 2018c).

For the purpose of detection, the gravitational waveform from the inspiral phase of a BNS coalescence is well modeled and matched filtering can be used to search for signals (Lindblom et al. 2008; Buonanno et al. 2009; Brown et al. 2012; Read et al. 2013; Abbott et al. 2016d; Harry et al. 2016). For systems containing black holes, or in which the component spin is significant, uncertainties in the waveform model can reduce the sensitivity of the search (Nitz et al. 2013; Harry et al. 2014; Taracchini et al. 2014; Pan et al. 2014; Dal Canton et al. 2015; Schmidt et al. 2015; Khan et al. 2016; Bustillo et al. 2017).

Searches for unmodeled transients make few assumptions on the signal morphology, using time–frequency decompositions to identify statistically significant excess-power transients in the data. The search for these transients focuses mainly on short-duration signals ($\lesssim 1$ s), but is also used for much longer signals (Abbott et al. 2019a). Their astrophysical targets include core-collapse supernovae, magnetar flares, BNS post-merger remnants, and as-yet-unknown systems (e.g., Klimenko et al. 2008; Sutton et al. 2010; Chassande-Mottin et al. 2010; Thrane et al. 2011; Adams et al. 2013; Thrane and Coughlin 2013; Cornish and Littenberg 2015; Thrane et al. 2015; Kanner et al. 2016). Expected detection rates for these transient sources are lower and/or less well constrained than CBCs. The burst search is complementary to the CBC search for BBH coalescences. It spans a larger parameter space with good efficiency to search for non-standard-BBHs, possible non-GR events, BBHs with eccentricity larger than 0.2, high-mass BBH systems, and intermediate mass black hole binaries (IMBHBs; Abadie et al. 2012e; Aasi et al. 2014c; Abbott et al. 2017l, 2019e, f). The search for short-duration gravitational-wave transients includes cosmic string cusps for which the waveform is well-modeled, and a matched-filter search is performed (Abbott et al. 2018b, 2019b).

During the observing runs, CBC and unmodeled searches are carried out in *near real-time* to rapidly identify event candidates and deliver prompt notice of potential GW transients enabling follow-up observations in the electromagnetic spectrum. Increased detection confidence, improved sky localization, identification of a host galaxy, and the source redshift are just some of the benefits of joint GW–electromagnetic observations. Here, we focus on two points of particular relevance for the rapid detection of GW transients and for the follow-up of candidate GW events: the GW signal significance and the source localization afforded by a GW detector network.

## 3.1 Detection and false alarm rates

Detection pipelines search the data looking for signal-like features. Candidate triggers flagged by a pipeline are assigned a detection statistic to quantify how signal-like they are. For CBC searches, this involves matching a bank of waveform templates (Sathyaprakash and Dhurandhar 1991; Owen 1996; Owen and Sathyaprakash 1999; Babak et al. 2006; Cokelaer 2007; Prix 2007; Harry et al. 2009; Ajith et al. 2014; Brown et al. 2012; Capano et al. 2016; Dal Canton and Harry 2017) to the data (Abbott et al. 2016d, b); for unmodeled searches, requirements on waveform morphology are relaxed, but coherence of the signal in multiple detectors is required (Abbott et al. 2016j, 2017b). A detection statistic is used to rank candidates; we assess significance by comparing results with those from an estimated background distribution of noise triggers. It is difficult to theoretically model the behaviour of non-Gaussian noise, and therefore the distribution must be estimated from the data (Abadie et al. 2010a, 2012a; Babak et al. 2013; Abbott et al. 2016a, b, d, j, 2017b; Capano et al. 2017; Messick et al. 2017; Nitz et al. 2017). From the background noise distribution we can map a value of the detection statistic to a false alarm rate (FAR), the expected rate of triggers with detection statistics equal to or greater than that value, assuming that the data contain no signals. While each pipeline has its own detection statistic, they all compute a FAR. The FAR, combined with the observation time, may then be used to calculate a $p$ value, the probability of there being at least one noise trigger with a FAR this low or lower in the observed time. The smaller the FAR or $p$ value of a trigger, the more significant it is, and the more likely that it is of astrophysical origin.

The $p$ value is distinct from the probability that a trigger is a real astrophysical GW signal, which we indicate as $p_{\mathrm{astro}}$. The $p$ value assumes that the data contain no signals, whereas the probability of there being a GW must include the hypothesis that there is an astrophysical signal. Thus, to calculate $p_{\mathrm{astro}}$ requires an extra layer of inference, folding in both our knowledge of trigger distribution, assumptions about signal distribution (such as that sources are uniformly distributed in volume), and knowledge and assumptions about merger rate per unit volume for each class of sources. A method to evaluate $p_{\mathrm{astro}}$ is described in Abbott et al. (2016b, m, n, 2018c) and Kapadia et al. (2020). The $p_{\mathrm{astro}}$ is given in the public GW alerts (see Sect. 4). Details on how it is evaluated in low-latency are given in the the LIGO/Virgo Public Alerts User Guide.[10]

The rate of noise triggers above a given detection statistic depends critically upon the data quality of the advanced detectors; non-stationary transients or *glitches* (Aasi et al. 2012, 2015b; Abbott et al. 2016c; Dal Canton et al. 2014a) produce an elevated background of loud triggers. Over 200, 000 auxiliary channels record data on instrumental and environmental conditions (Effler et al. 2015; Abbott et al. 2016c). These channels act as witnesses to disturbances that may couple into the GW channel (Berger 2018; Walker et al. 2018; Covas et al. 2018; Zevin et al. 2017). However, it is not always possible to identify what produces certain glitches. An intensive study of the quality of the data is used to veto stretches ranging from seconds to hours in duration (Nuttall et al. 2015). When a significant problem with the data is identified or a known instrumental issue affects the searches' background, the contaminated data are removed from the analysis data set. Our experience to date is that this removes a small percentage of the data. For CBC searches, the waveforms are well modeled, and signal consistency tests reduce the background significantly (Allen 2005; Cannon et al. 2015; Usman et al. 2016). For burst sources which are not well modeled, or which spend only a short time in the detectors' sensitive band, it is more difficult to distinguish between the signal and a glitch. Consequently a reduction of the FAR threshold comes at a higher cost in terms of reduced detection efficiency.

Search pipelines are run both online, analysing data as soon as they are available in order to provide low-latency alerts of interesting triggers, and offline, taking advantage of improved calibration of the data and additional information regarding data quality. In Fig. 3, we show the results of the offline transient searches performed during O1 and O2. In each plot we show the observed distribution of events as a function of inverse false alarm rate (IFAR), as well as the expected background for the analysis. The FAR of the eleven confident gravitational wave detections are reported in the GWTC-1 catalog (Abbott et al. 2018c) and (Abbott et al. 2019b). Full strain data from O1 and O2, as well as auxiliary data for GW events and software to analyze GW data, are publicly available from the LIGO and Virgo Gravitational Wave Open Science Center[11] (Vallisneri et al. 2015). Publication of a GW event is accompanied by the release of strain data around the time of that event. Data from O3 and subsequent runs will be available at the same location (Anderson and Williams 2017).

[10] The User Guide is available at emfollow.docs.ligo.org/userguide/.

[11] www.gw-openscience.org.

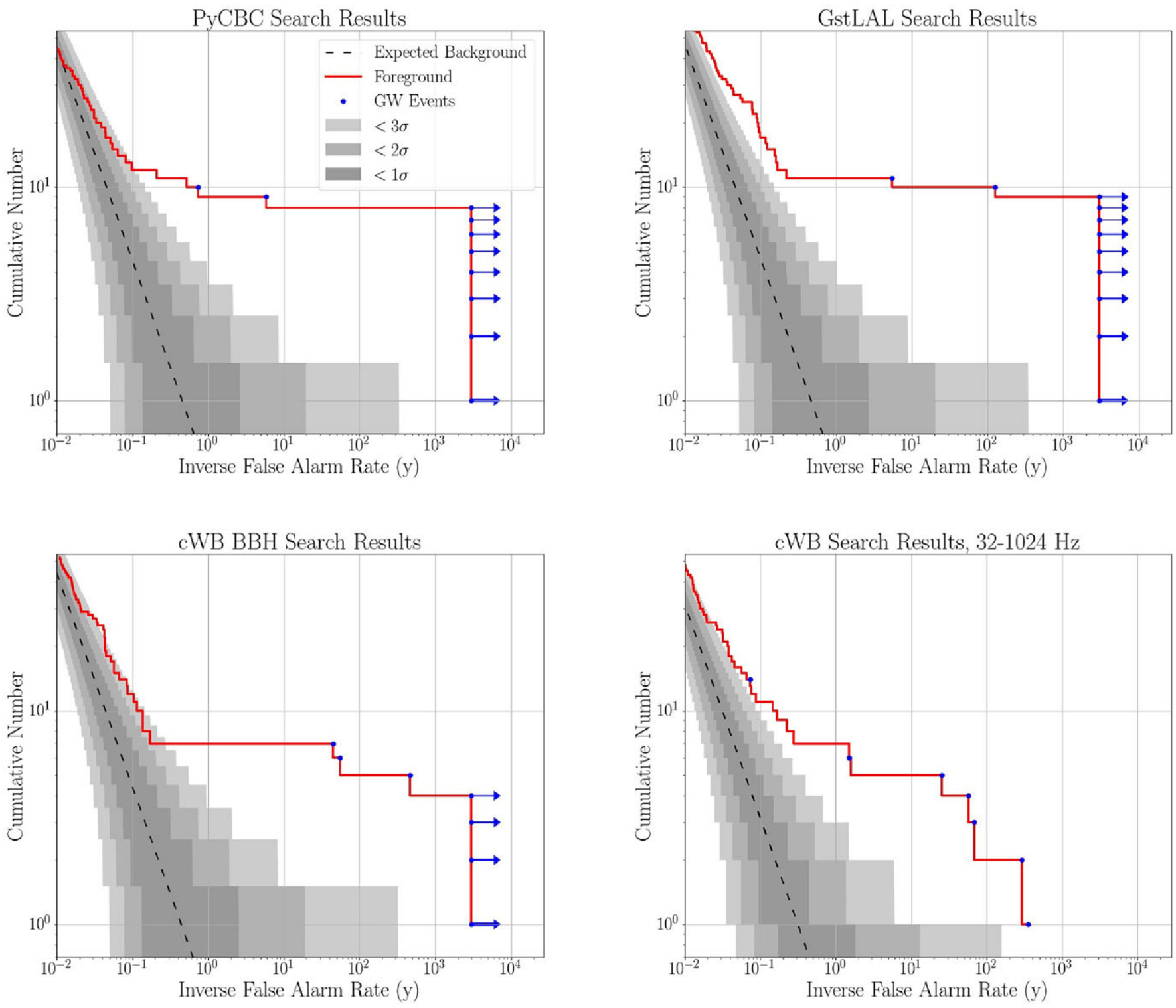

**Fig. 3** Cumulative histograms of triggers obtained by the offline searches plotted versus the IFAR. The top panel shows results for the matched-filter searches; on the left the PyCBC (Dal Canton et al. 2014b; Usman et al. 2016) search pipeline, and on the right the GstLAL (Cannon et al. 2012; Privitera et al. 2014; Messick et al. 2017; Sachdev et al. 2019) search pipeline. The bottom panels show unmodeled searches performed by the cWB (Klimenko et al. 2008, 2016) pipeline; on the left looking for stellar-mass BBHs mergers, and on the right for generic transients. The dashed lines show the expected background, given the analysis time. Shaded regions denote the sigma uncertainty bounds for the Poisson statistic. The blue dots are the confident GW events found by each search. Any events with a measured or bounded inverse false alarm rate greater than 3000 years are shown with a right pointing arrow. The values of the FARs of the confident events can be found in Abbott et al. (2017b, 2018c, 2019b)

## 3.2 Localization

Following the detection of a GW transient, posterior probability distributions for the position are constructed following a Bayesian framework (Veitch et al. 2015; Cornish and Littenberg 2015; Singer and Price 2016; Abbott et al. 2016k), with information for the sky localization coming from the time of arrival, plus the phase and amplitude of the GW signal.

An intuitive understanding of localization can be gained by considering triangulation using the observed time delays between sites (Fairhurst 2009, 2011). The effective single-site timing accuracy is approximately

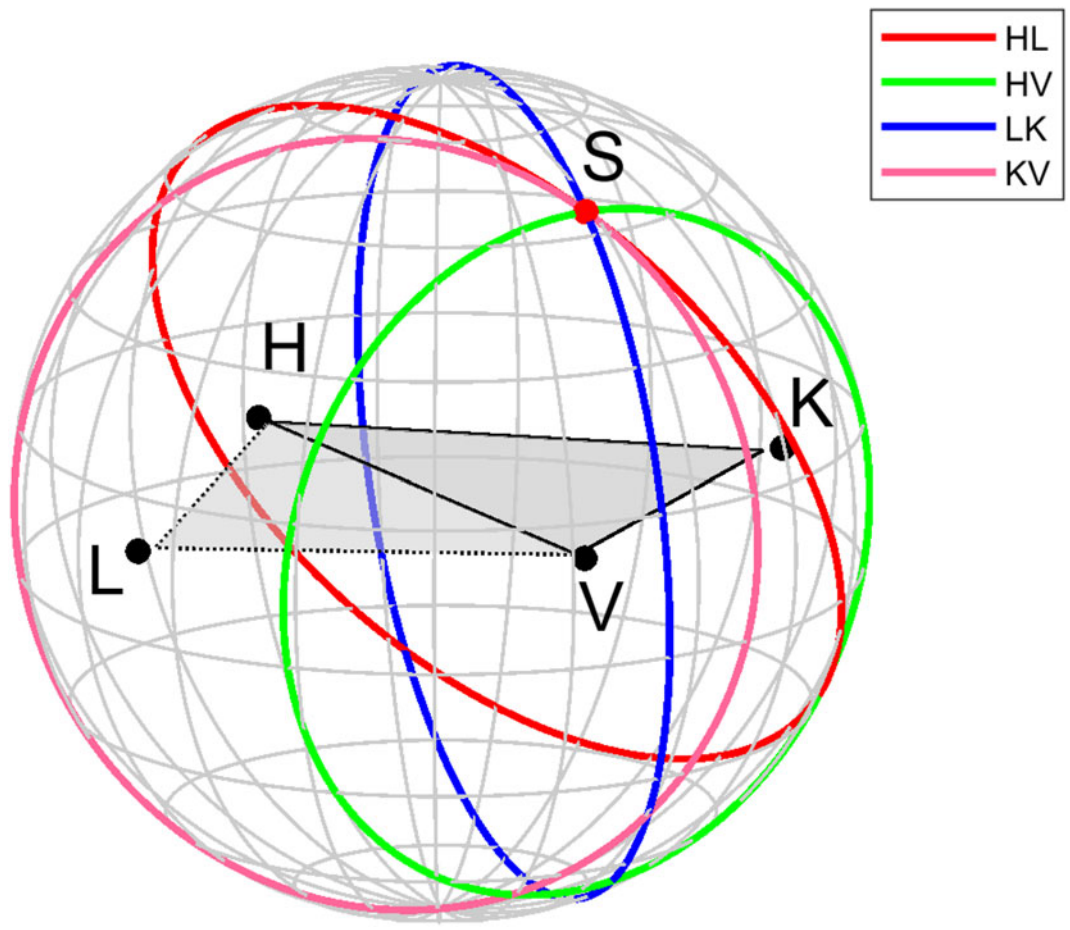

**Fig. 4** Source localization by timing triangulation for the aLIGO–AdV–KAGRA network. The locations of the four detectors are indicated by black dots, with LIGO Hanford labeled H, LIGO Livingston as L, Virgo as V and KAGRA as K. The locus of constant time delay (with associated timing uncertainty) between two detectors forms an annulus on the sky concentric about the baseline between the two sites (labeled by the two detectors). For clarity we omit the HK and LV combinations. For four or more detectors there is a unique intersection region, *S*. Image adapted from Chatterji et al. (2006)

$$\sigma_t = \frac{1}{2\pi\rho\sigma_f}, \qquad (2)$$

where $\rho$ is the SNR in the given detector and $\sigma_f$ is the effective bandwidth of the signal in the detector, typically of order 100 Hz. Thus a typical timing accuracy is on the order of $10^{-4}$ s (about 1/100 of the typical light travel time between sites, which is of order 10 ms). This sets the localization scale. The simple model of Eq. (2) ignores many other relevant issues such as information from the signal amplitudes and phases across the detector network, uncertainty in the emitted gravitational waveform, and instrumental calibration accuracies. The source sky location of CBC signals is currently evaluated by introducing the requirement of phase and amplitude consistency between detectors (Grover et al. 2014; Fairhurst 2017). A Bayesian inference algorithm constructs posterior probability distributions for the system parameters—location, mass, distance, orientation, etc.—by matching GW models to the detector strain (Cutler and Flanagan 1994; Röver et al. 2007a, b; Fairhurst 2009, 2017; Vitale and Zanolin 2011; Vitale et al. 2012; Nissanke et al. 2011, 2013; Veitch et al. 2012; Jaranowski and Królak 2012; Aasi et al. 2013b; Singer et al. 2014; Berry et al. 2015; Singer and Price 2016; Abbott et al. 2017c).

Source localization using only timing for a two-site network yields an annulus on the sky; see Fig. 4. Adding the signal amplitude and phase (and also precession effects) resolve this to only parts of the annulus. However, even then sources will be localized to regions of hundreds to thousands of square degrees (Singer et al. 2014; Berry et al. 2015).

For three detectors, the time delays restrict the source to two sky regions which are mirror images with respect to the plane passing through the three sites. Requiring consistent amplitudes and phase in all the detectors typically eliminates one of these regions (Fairhurst 2017). This typically yields regions with areas of several tens to hundreds of square degrees. If there is a significant difference in sensitivity between detectors, the source is less well localized and we may be left with the majority of the annulus on the sky determined by the two most sensitive detectors. With four or more detectors, timing information alone is sufficient to localize to a single sky region, and the additional baselines help to localize within regions smaller than ten square degrees for some signals.

From Eq. (2), it follows that the *linear* size of the localization ellipse scales inversely with the SNR of the signal and the frequency bandwidth of the signal in the detector (Berry et al. 2015). For GWs that sweep across the band of the detector, such as CBC signals, the effective bandwidth is $\sim 100\,\mathrm{Hz}$. Higher mass CBC systems merge at lower frequencies and so have a smaller effective bandwidth. For burst signals, the bandwidth $\sigma_f$ depends on the specific signal. For example, GWs emitted by various processes in core-collapse supernovae are anticipated to have relatively large bandwidths, between 150 and 500 Hz (Dimmelmeier et al. 2008; Ott 2009; Yakunin et al. 2010; Ott et al. 2011). By contrast, the sky localization region for narrowband burst signals may consist of multiple disconnected regions and exhibit fringing features; see, for example, Klimenko et al. (2011), Abadie et al. (2012c) and Essick et al. (2015).

The sky localization of GW events confidently detected during O1 and O2 and sent in low-latency is shown in the top plot of Fig. 5. The refined sky localization obtained offline by the parameter estimation analysis is shown in the bottom plot of the same figure. The offline analyses exploit refined instrumental calibration, noise subtraction, updated estimates of the amplitude power spectral density, and extended template banks (Abbott et al. 2018c, 2019d). The plots show that even if the posterior probability is primarily distributed along a ring, the ring is broken into disconnected components determined by the sensitivity of the individual detectors. The events detected by the two LIGO interferometers show the expected trend of the sky area to scale inversely with the square of the SNR (Abbott et al. 2018c). Five of the 11 confident events were observed with the three-site HLV network (see Table 3). The Virgo data were used to estimate the low-latency sky localization for two events (GW170814 and GW170817). With the contribution from the third detector we were able to significantly shrink the localization to areas covering a few tens of square degrees (see GW170814, GW170817, GW170718).

In addition to localizing sources on the sky, it is possible to provide distance estimates for CBC signals since the waveform amplitude is inversely proportional to the luminosity distance (Veitch et al. 2015; Abbott et al. 2016k). Uncertainty in distance measurement is dominated by the degeneracy with the inclination of the binary, which also determines the signal amplitude (Cutler and Flanagan 1994; Röver et al. 2007a; Nissanke et al. 2010; Aasi et al. 2013b). The degeneracy could be broken by observing with more non-co-aligned detectors (Veitch et al. 2012; Rodriguez et al. 2014), or if precession of the orbital plane is observed (Vecchio

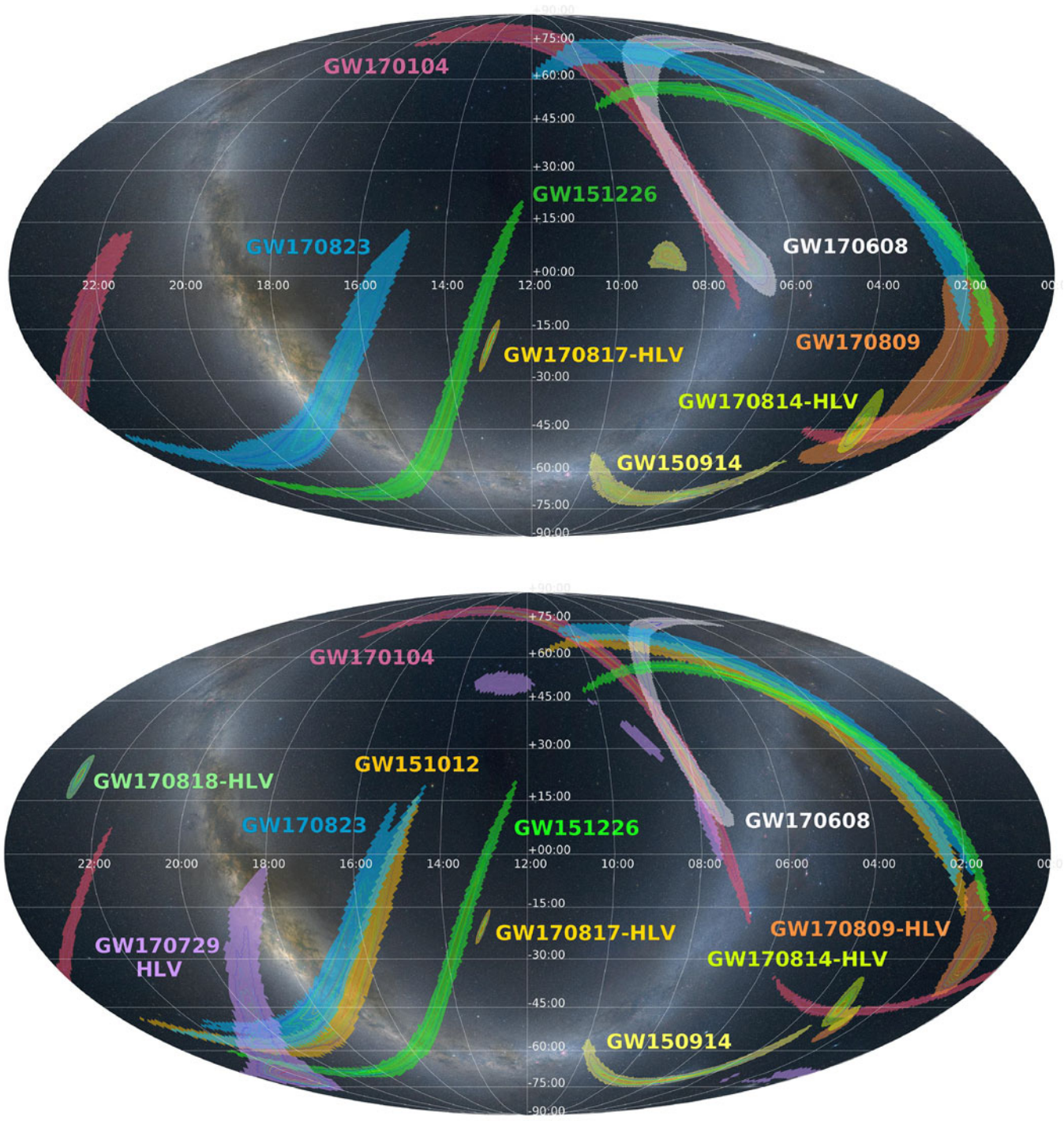

**Fig. 5** Sky locations of GW events confidently detected in O1 and O2. Top panel: initial sky location released in low-latency to the astronomers (Abbott et al. 2016h; LIGO Scientific Collaboration and Virgo Collaboration 2015; Abbott et al. 2019d). Bottom panel: refined sky location including updated calibration and final choice of waveform models (Abbott et al. 2018c). Three events (GW151012, GW170729, GW170818) among the 11 confidetent detections were identified offline, and were not shared in low-latency. The shaded areas enclose the 90% credible regions of the posterior probability sky areas in a Mollweide projection. The inner lines enclose regions starting from the 10% credible area with the color scheme changing with every 10% increase in confidence level. The localization is shown in equatorial coordinates (right ascension in hours, and declination in degrees). The HLV label indicates events for which both the LIGO and Virgo data were used to estimate the sky location

2004; van der Sluys et al. 2008; Vitale et al. 2014), but this is not expected for slowly spinning BNS (Farr et al. 2016). Distance information can further aid the hunt for counterparts, particularly if the localization can be used together with galaxy catalogs (Abadie et al. 2012c; Nissanke et al. 2013; Hanna et al. 2014; Fan et al. 2014; Blackburn et al. 2015; Singer et al. 2016a; Del Pozzo et al. 2018). Table 3 reports the low-latency and refined estimates for the luminosity distance and the sky localization (90% credible region) of the eleven confident signals detected during O1 and O2.[12]

Some GW searches are triggered by electromagnetic observations, and in these cases initial localization information is typically available a priori. For example, in

[12] The initial sky maps are available from dcc.ligo.org/public/0160/P1900170/001/O1_O2_LowLatency_Skymaps.zip, and the refined sky maps from dcc.ligo.org/LIGO-P1800381/public, respectively.

**Table 3** Luminosity distance $d_L$ and sky localization $\Delta\Omega$ for the eleven confident signals detected during O1 and O2

| Event | Low-latency analysis | | | Refined analysis | | |
|---|---|---|---|---|---|---|
| | $d_L$(Mpc) | $\Delta\Omega(\mathrm{deg}^2)$ | IFOs | $d_L$(Mpc) | $\Delta\Omega(\mathrm{deg}^2)$ | IFOs |
| GW150914 | – | 307 | HL | $440^{+150}_{-170}$ | 182 | HL |
| GW151012 | – | – | – | $1080^{+550}_{-490}$ | 1523 | HL |
| GW151226 | – | 1337 | HL | $490^{+180}_{-190}$ | 1033 | HL |
| GW170104 | $730^{+340}_{-320}$ | 1632 | HL | $990^{+440}_{-430}$ | 921 | HL |
| GW170608 | $310^{+200}_{-120}$ | 864 | HL | $320^{+120}_{-110}$ | 392 | HL |
| GW170729 | – | – | – | $2840^{+1400}_{-1360}$ | 1041 | HLV |
| GW170809 | $1080^{+520}_{-470}$ | 1155 | HL | $1030^{+320}_{-390}$ | 308 | HLV |
| GW170814 | $480^{+190}_{-170}$ | 97 | HLV | $600^{+150}_{-220}$ | 87 | HLV |
| GW170817 | $40^{+10}_{-10}$ | 31 | HLV | $40^{+7}_{-15}$ | 16 | HLV |
| GW170818 | – | – | – | $1060^{+420}_{-380}$ | 39 | HLV |
| GW170823 | $1380^{+700}_{-670}$ | 2145 | HL | $1940^{+970}_{-900}$ | 1666 | HL |

The distances are given as median value with 90% credible intervals, and the sky localizations as the 90% credible areas. For event detected in low-latency columns 2 and 3 show the initial source parameters, which were obtained by the on-line analysis (Abbott et al. 2019d). Columns 5 and 6 show the source parameter obtained by the offline refined analysis (Abbott et al. 2018c). The IFOs columns indicate the detector data used for the parameter estimation. All the initial sky maps were produced by BAYESTAR, except GW150914, which was detected in low-latency by the unmodeled search cWB (Abbott et al. 2016h). The final refined sky maps are produced by LALINFERENCE. Details about localization pipelines are given in Sects. 3.2.1 and 3.2.2. GW151012, GW170729, GW170818 were identified offline, and were not shared in low-latency. The distance of GW150914 and of GW151226 were not shared in low-latency following the policy applied in O1. In contrast to the median luminosity distances listed here, the sky map headers(see footnote 12) list the posterior mean and standard deviation

GW searches triggered by gamma-ray bursts (Abadie et al. 2012d; Aasi et al. 2014b, d; Abbott et al. 2017k), the triggering space-based telescope provides a localization. The rapid identification of a GW counterpart to such a trigger will prompt longer and deeper follow-up in different wavelengths that may not always be done in response to gamma-ray bursts (cf. Abbott et al. 2017j). This is particularly important for gamma-ray bursts with larger sky localization uncertainties, such as those reported by *Fermi*-GBM (Meegan et al. 2009), which are not followed up as frequently as the bursts reported by the *Neil Gehrels Swift Observatory* (Gehrels et al. 2004) or *Fermi*-LAT (Atwood et al. 2009), which provide good sky localization. In the case of GW170817, the LIGO–Virgo localization was tighter than the localization from *Fermi*-GBM and *INTEGRAL* (Abbott et al. 2017e; Goldstein et al. 2017a; Savchenko et al. 2017a) and showed that the source was nearby ($40^{+8}_{-14}$ Mpc; Abbott et al. 2017i), making it a prime target for further follow-up. Other possible targets for externally-triggered GW searches are electromagnetic or neutrino emission from soft-gamma ray repeaters and pulsar glitches (Abadie et al. 2008, 2011b; Lasky 2015; Abbott et al. 2019g).

All GW data are stored permanently, so that it is possible to perform retroactive analyses at any time.

### 3.2.1 Localization for compact binary coalescences

Providing prompt localizations for GW signals helps to maximise the chance that electromagnetic observatories can find a counterpart. Localizations are produced at several different latencies, with updates coming from more computationally expensive algorithms that refine our understanding of the source.

For CBC signals, rapid localization is performed using BAYESTAR (Singer and Price 2016), a Bayesian parameter-estimation code that computes source location using output from the detection pipeline. BAYESTAR produces sky localizations (as in Fig. 5, top plot) with latencies of only a few seconds. It also provides distance estimates (Singer et al. 2016a). These are communicated as an additional component of the sky localization (3D sky map): for each line of sight, the distance posterior probability is approximated as a Gaussian multiplied by the distance squared (Singer et al. 2016a, b).[13] Results from BAYESTAR are shared in low latency for prompt electromagnetic/neutrino follow-up.

At higher latency, the CBC parameter estimation is performed using the Bayesian inference algorithms of LALINFERENCE (Veitch et al. 2015), which constructs posterior probability distributions for the system parameters, and not just location like BAYESTAR. Computing waveforms for a large number of source parameters is computationally expensive; this expense increases as the detectors' low-frequency sensitivity improves and waveforms must be computed down to lower frequencies. The quickest LALINFERENCE binary system coalescence follow-up is computed using waveforms that do not include the full effects of component spins (Singer et al. 2014; Berry et al. 2015; Abbott et al. 2017f). Localizations are reported with latency of hours to several days. Parameter estimation is then performed using more accurate waveform approximants, those that include full effects of spin precession and the effects of tidal distortions of neutron stars (Farr et al. 2016; Abbott et al. 2016g, 2017f, i). Provided that BNSs are slowly spinning (Mandel and O'Shaughnessy 2010), the restrictions on the spins should cause negligible difference between the mid-latency LALINFERENCE and the high-latency fully spinning LALINFERENCE localizations (Farr et al. 2016). Methods of reducing the computational cost are actively being investigated (e.g., Canizares et al. 2013; Pürrer 2014; Canizares et al. 2015; Smith et al. 2016; Vinciguerra et al. 2017). Parameter estimation through Bayesian inference is an active field of research and new algorithms are currently being considered (Ashton et al. 2019).

Differences between the BAYESTAR and LALINFERENCE localizations are expected to be negligible, except in the case of strong precession of the binary system (Farr et al. 2016), because BAYESTAR uses the maximum likelihood template from the low-latency detection pipelines which do not currently include precession. Differences among the low- and mid-latency sky maps are possible as improvements are made in the handling of data calibration and the characterisation of the noise. Significant shifts and shape changes of the sky maps, such as for GW170814 (Abbott et al. 2019d), are expected only in the case of problems in the data calibration, data quality or glitch treatment.[14]

[13] A data release of example three-dimension localizations in this format, constructed using results from BAYESTAR and LALINFERENCE for BNS signals, is available from dcc.ligo.org/P1500071/public/html.

Figure 6 shows the expectations for the sky localization of astrophysically motivated populations of BNS, NSBH, and BBH signals during O3 and O4. For O3, we consider two scenarios; the HLV network, and the HLVK network. For O4, we consider only the HLVK network. We assume a source to be detected if it has SNR larger than 4 in at least two detectors and a network SNR larger than 12. This is a conservative threshold, considering that some of the GW events confidently detected in O1 and O2 have a network SNR smaller than 12 (Abbott et al. 2018c). It is also larger than the SNR threshold (of about 8.5) corresponding to the FAR used to release GW candidate alerts associated with binary systems of compact objects during O3 (see Sect. 4). We use: 1) a population of BNSs with component masses drawn from a Gaussian distribution with mean 1.33 and standard deviation 0.09, and spins aligned or anti-aligned with uniformly distributed magnitudes smaller than 0.05; 2) a population of BBHs with the primary masses distributed as a power-law with index of $\alpha = -2.3$, mass range 5–$50\,M_{\odot}$, and spins aligned or anti-aligned with uniformly distributed magnitudes smaller than 0.99, and 3) a NSBH population with the mass and spin distributions described for the BNSs and BBHs. The merger rate density is assumed constant in the comoving frame and source-frame time. The results of our simulation are quantified using the GW signal sky-localization area, luminosity distance, and comoving volume. Sky-localization area (volume) is given as the 90% credible region, defined as the smallest area (volume) enclosing 90% of the total posterior probability. This coresponds to the area (volume) of the sky that must be covered to have a 90% chance of including the source.

During O3 the expected four-detector localizations are only slightly better than the three-detector ones (the median 90% credible area is reduced by about 30%). This is due to the limited sensitivity of KAGRA with respect to the other detectors, which only significantly improves the localization of loud signals. A large improvement of the localization capability (area and volume) is shown for O4, where the expanded network of detectors is accompanied by higher sensitivities. The 90% credible regions for the area and the volume are shown in Table 5 and discussed further in Sects. 5.1 and 5.2. The effects on the sky localization of BNS and NSBH signals from assuming different astrophysical mass and spin distributions is discussed in Pankow et al. (2019).

[14] The O3 run showed differences among the BAYESTAR and LALINFERENCE which are quantified and justified in Sect. 4.3.

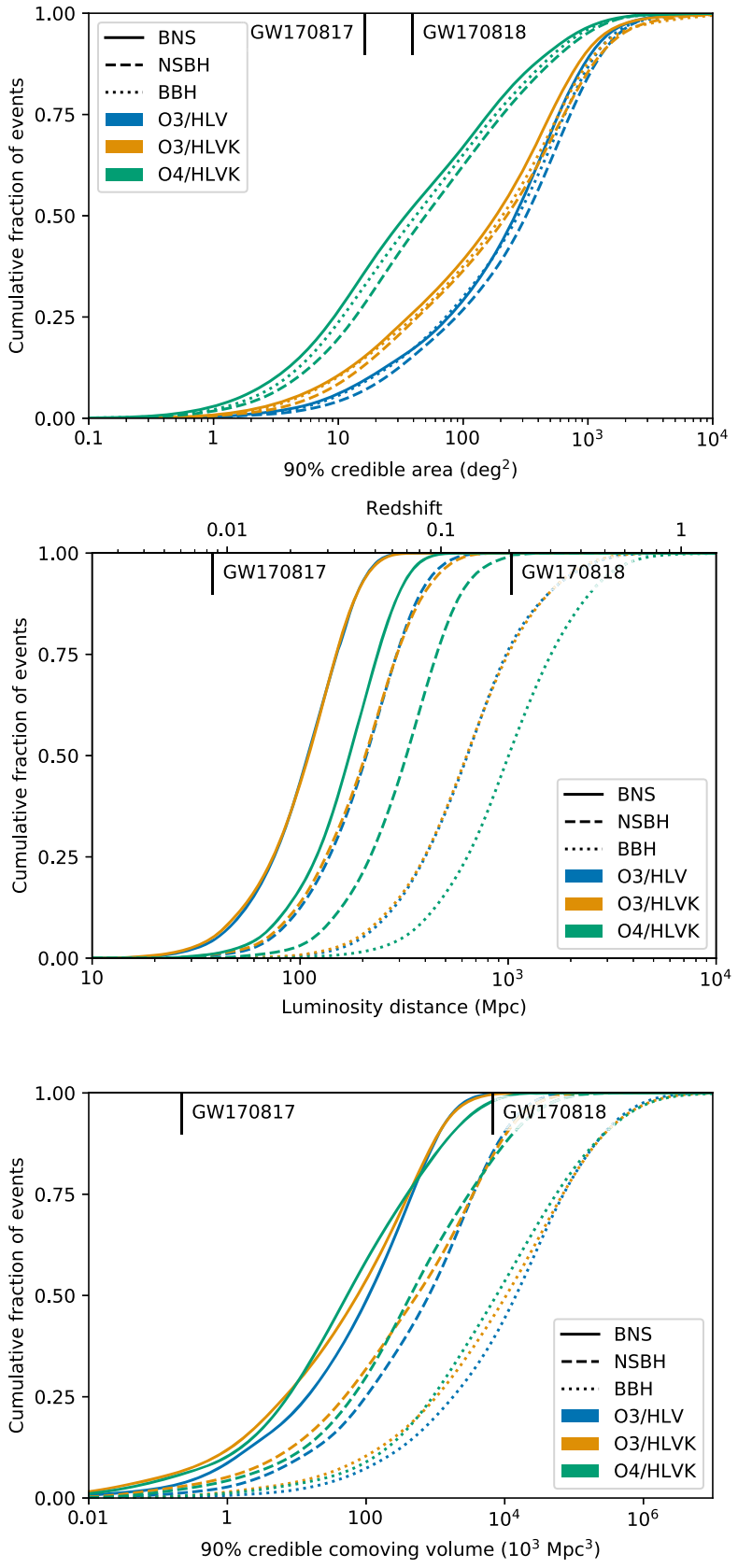

**Fig. 6** Anticipated GW sky localization for CBC signals during the third and fourth runs (for O3, see Sect. 5.1 and for O4, see Sect. 5.2). For O3, the detector sensitivities were taken to be representative of the first 3 months of observations for aLIGO Hanford and Livingston, and AdV, and the highest expected O3 sensitivity for KAGRA (see Fig. 1). For O4, the detector sensitivities were taken to be the target sensitivities for aLIGO and AdV, and the mid of the interval expected for KAGRA during O4. Top: The plot shows the cumulative fractions of events with sky-localization area smaller than the abscissa value. Central: The plot shows the cumulative fractions of events with luminosity distance smaller than the abscissa value. Bottom: The plot shows the cumulative fractions of events with comoving volume smaller than the abscissa value. Sky-localization area (comoving volume) is given as the 90% credible region, the smallest area (comoving volume) enclosing 90% of the total posterior probability. Results are obtained using the low-latency BAYESTAR pipeline (Singer and Price 2016). The simulation accounts for an independent 70% duty cycle for each detector, and the different sensitivity of each sub-network or network of detectors. For O3, all the combinations of sub-networks of two operating detectors and the three detector network (HLV) are included in the blue lines. All the combinations of sub-networks of two and three operating detectors, and the four detector network (HLVK) are included in the orange lines for O3 and in the green lines for O4. The O3 HLV and the O3 HLVK curves in the central panel are very similar due to the modest contribution by KAGRA to the network SNR. Solid lines represent BNSs, dashed lines NSBHs, dotted lines BBHs. As a comparison, the plots show the area, distance and volume of GW170817 and GW170818, which are the best localized BNS and BBH signals during O1 and O2

LALInference has the ability to include the effects of the detectors' calibration uncertainty on parameter estimation (Abbott et al. 2016b, k). Initial results for GW150914 assumed a calibration uncertainty of 10% for the amplitude of the GW strain and 10 deg for its phase (Abbott et al. 2017c). Incorporating this calibration uncertainty into the analysis, the 90% credible area was 610 $\deg^2$ (Abbott et al. 2016k). By the end of O1, the calibration uncertainty had been improved, such that the 90% credible area was 230 $\deg^2$ (Abbott et al. 2016b). If the detectors were assumed to be perfectly calibrated, such that calibration uncertainty could be ignored, the 90% credible area would be 150 $\deg^2$. The sky localization is particularly sensitive to calibration uncertainty, while distance is less affected. For GW150914, the initial distance estimate was $410^{+160}_{-180}$ Mpc (Abbott et al. 2016k), the estimate at the end of the run was $420^{+150}_{-180}$ Mpc, and the equivalent result without calibration uncertainty was $420^{+140}_{-170}$ Mpc (Abbott et al. 2016b). The effects of calibration uncertainty depend upon the signal's SNR, bandwidth and the position of the source relative to the detectors. For example, for GW151226, GW151012 and GW170104, there is negligible difference between the sky areas or distances with and without final calibration uncertainties (Abbott et al. 2016b, 2017f).

The targets for O3 on the calibration uncertainties are $< 3\%$ for the amplitude of the GW strain and $< 2$ deg for its phase at 68% confidence interval, from 20 to 1024 Hz. This includes a site-to-site timing uncertainty of $\sim 1\,\mu$s. This information is folded into the parameter estimation of CBC candidate events over which the uncertainties are marginalized. The current techniques for this marginalization are discussed in Farr et al. (2015).

### 3.2.2 Localization for unmodeled signals

Sky localizations are also produced for unmodeled triggers and distributed for follow up. The lowest latency sky localizations are produced as part of the coherent Wave Burst (cWB) detection pipeline (Klimenko et al. 2008, 2016). Sky localizations are produced using a constrained likelihood algorithm that coherently combines data from all the detectors. The cWB sky localizations are calculated with a latency of a few minutes.

Following detection, an unmodeled burst signal is analyzed by parameter-estimation codes: LALInferenceBurst (LIB), a stochastic sampling algorithm similar to the LALInference code used to reconstruct CBC signals (Veitch et al. 2015), and BayesWave, a reversible jump Markov-chain Monte Carlo algorithm that models both signals and glitches (Cornish and Littenberg 2015). LIB uses sine–Gaussian waveforms (in place of the CBC templates used by LALInference), and can produce sky localizations in a few hours. BayesWave uses a variable number of sine–Gaussian wavelets to model the signal and the glitches while also fitting for the noise spectrum using BayesLine (Littenberg and Cornish 2015); it produces sky localizations with a latency of minutes.

The sky-localization performance of unmodeled algorithms depends upon the type of signal. Studies of burst localization using BayesWave in the first year of the advanced-detector era, and using cWB and LIB in the first 2 years have been completed in Bécsy et al. (2017) and Essick et al. (2015), respectively. These works show results for a variety of waveform morphologies that could be detected in a burst search (Abadie et al. 2012c): Gaussian, sine-Gaussian, broadband white-noise and BBH waveforms.

We present sky localization results obtained by cWB for two astrophysically motivated populations, which are expected to emit signals detectable by burst searches: the mergers of BBHs and the mergers of IMBHBs. We assume a population of BBHs with total mass less than $100\,M_{\odot}$, distribution of the primary mass uniform in the log, component masses in the 5–$50\,M_{\odot}$ range, and isotropic distribution of the spin. The population of IMBHBs is composed of black holes of individual mass $100\,M_{\odot}$, and with spins aligned with the binary orbital angular momentum. To search for these signals cWB identifies regions of excess power in the time-frequency representation of the gravitational strain. The search pattern is optimized with a different selection of pixels tuned for BBHs and IMBHBs, respectively. The cWB searches optimized for BBH and IMBHB currently run in low-latency together with the standard cWB. Figure 7 shows the sky localization area for BBHs (*Left plots*) and IMBHBs (*Right plots*) for the LIGO network (HL), for the LIGO and Virgo network (HLV), and the LIGO, Virgo and KAGRA network (HLVK)[15] during O3 (*Top plots*) and O4 (*Bottom plots*). The median BBH sky-localization obtained with the unmodeled search is 490 (220) $\mathrm{deg}^2$ with three (four) detectors in O3. It reduces to about 90 $\mathrm{deg}^2$ in O4 with four sensitive detectors. The IMBHB sky-localization is larger; 730 (510) $\mathrm{deg}^2$ with HLV (HLVK) in O3, and 360 $\mathrm{deg}^2$ with HLVK in O4. The anticipated ranges of the cWB searches for BBH mergers and IMBHB mergers during O3 and O4 are reported in Table 4. The unmodeled searches for BBHs and IMBHBs are able to reach ranges up to the gigaparsec scale.

### 3.3 The O1 and O2 follow-up program

During the first (O1) and second (O2) observing runs, GW candidate alerts were sent privately to groups of astronomers who signed an MOU with the LIGO Scientific Collaboration (LSC) and Virgo collaborations. At the end of O2, the follow-up program included 95 groups, with capabilities to search for electromagnetic counterparts from very high-energy to the radio band, and to search for neutrino counterparts. The low-latency identification and validation of GW signal candidates, and the distribution of alerts is detailed in Abbott et al. (2019d). Only candidates with a FAR below a threshold of once per 2 months were selected to trigger the search for counterparts. Properties of the GW candidates were distributed using the Gamma-ray Coordinates Network (GCN) system,[16] widely used in the astronomical community for the multiwavelength follow-up of gamma-ray bursts. The GCNs included event time, sky localization probability map, and the estimated FARs. For compact binary merger candidates, they also included volume localization (3D sky map), probability of the system to contain a neutron star and probability to be electromagnetically bright (based on the estimate of the baryon mass left outside the merger remnant, Foucart 2012; Pannarale and Ohme 2014).

[15] In contrast to the CBC simulation, the burst sky localization simulation uses a specific network (HL, HLV, HLVK) without including sub-networks.

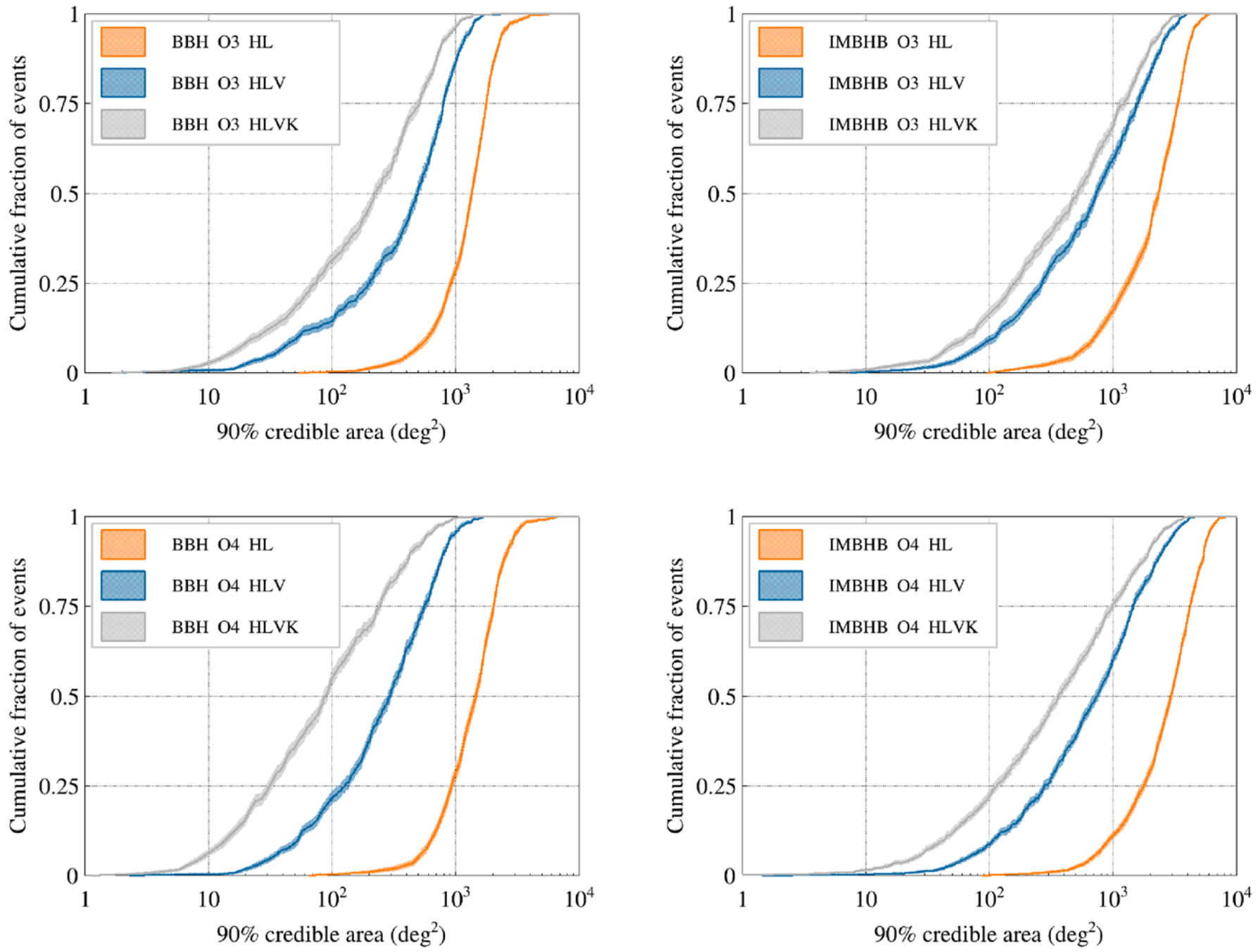

**Fig. 7** Simulated sky localization for unmodeled searches for mergers of BBHs and mergers of IMBHBs. The simulation uses a population of BBHs with the distribution of the primary mass uniform in the log, component masses in the $5–50\,M_\odot$ range and isotropic distribution of the spin. The population of IMBHBs is composed of black holes of individual mass $100\,M_\odot$, and with spins aligned with the binary orbital angular momentum. The plots show the cumulative fractions of events with 90% credible areas smaller than the abscissa value. The results obtained by the low-latency Coherent Wave Burst pipeline (Klimenko et al. 2005, 2008, 2016) for the third (Top plots—O3) and fourth observing runs (Bottom plots—O4) consider separately the HL, HLV and HLVK networks (without including sub-networks). These specific network configurations will be operating for a limited interval of time during the run. Assuming an instrument duty cycle of 70%, the HL network and HLV network would be operational 14% and 34% of the time during O3. Once KAGRA joins the observations, the HL, HLV, and HLVK networks will be operational 4%, 10%, and 24% of the time, respectively. The detection thresholds for cWB are set to 0.7 for the network correlation coefficient and 12 for the network SNR (see Abbott et al. 2018c). Shaded regions denote the 1-sigma uncertainty

[16] Details of the GCN are available from gcn.gsfc.nasa.gov.

**Table 4** Range of the cWB searches for merging BBHs with a total mass less than $100\,M_{\odot}$ and merging IMBHBs with component masses of $100\,M_{\odot}$

| Run | IFO net | BBH systems range (Mpc) | IMBHB systems range (Mpc) |
|---|---|---|---|
| O3 | LH | 700 | 2240 |
| O3 | LHV | 710 | 2290 |
| O3 | LHVK | 700 | 2280 |
| O4 | LH | 990 | 3070 |
| O4 | LHV | 1070 | 3250 |
| O4 | LHVK | 1060 | 3270 |

The range corresponds to the orientation-averaged spacetime volume surveyed per unit detector time. The range is given for the HL, HLV and HLVK networks for O3 and O4. The 1-sigma error on the range estimates is around 1–2 percent. While the inclusion of KAGRA improves the sky-localization, the detection efficiency remains the same (the HLV and HLVK ranges are consistent within the errors). The range evaluations are obtained using the simplified assumption of Gaussian noise, and the values can be considered as indicative of range expectations

Seventeen alerts were sent to the astronomers during O1 and O2. Among them seven signals are confident detections originating from BBHs (Abbott et al. 2016f, i, 2017f, g, h, 2018c) and one confident signal from a BNS, GW170817 (Abbott et al. 2017i). Four BBH mergers were detected in low-latency by the aLIGO interferometers, while three BBH mergers (GW170809, GW170814, GW170823), and the BNS merger GW170817 were observed with Advanced Virgo as part of the network of GW detectors. The inclusion of the third detector significantly improves the sky localization for the majority of these events (see e.g. Abbott et al. 2017h, 2018c), and consequently the efficiency of searches for electromagnetic counterparts.

For each GW trigger, tens of teams responded to the alert and operated ground- and space-based instruments spanning 19 orders of magnitude in electromagnetic wavelength (see e.g.; Abbott et al. 2016h; Cowperthwaite et al. 2016; Smartt et al. 2016; Racusin et al. 2017; Evans et al. 2016b; Palliyaguru et al. 2016; Abbott et al. 2017j, and references therein) The search for electromagnetic signatures of the GW source includes analysis of archival data around the time of the GW trigger, follow-up by covering the sky map or targeting the galaxies in the GW localization, and photometric and spectroscopic follow-up of the electromagnetic counterpart candidates by larger telescopes to remove contaminants and characterize the source. No firm electromagnetic counterpart has been found for any of the detected BBHs. A weak transient was found in *Fermi*-GBM data 0.4 s after GW150914 (Connaughton et al. 2016; Bagoly et al. 2016; Connaughton et al. 2018; Burns et al. 2019), and a weak signal was found in the AGILE-MCAL data 0.46 s before GW170104 (Verrecchia et al. 2017), but neither signal was confirmed by other satellites (Savchenko et al. 2016, 2017b; Tavani et al. 2016; Hurley et al. 2016; Goldstein et al. 2017b).

GW170817 was the first GW transient consistent with the coalescence of a BNS (Abbott et al. 2017i) and with the first firm electromagnetic counterpart (Abbott et al. 2017j). A prompt gamma-ray signal GRB 170817A (Goldstein et al. 2017a) was detected $\sim 1.7$ s after the merger time by *Fermi*-GBM, and later confirmed by INTEGRAL (Savchenko et al. 2017a). The three-detector GW localization led to the discovery of the bright transient AT 2017gfo by the One-Meter, Two-Hemisphere team with the 1-m Swope Telescope (Coulter et al. 2017), and confirmed by other teams within an hour (Soares-Santos et al. 2017; Valenti et al. 2017; Arcavi et al. 2017; Tanvir et al. 2017; Lipunov et al. 2017). Observations from the near infrared to the ultraviolet showed a transient thermal emission with a blue component fading within 2 days and a red-ward evolution in 1 week (e.g., Villar et al. 2017). An X-ray signal (Troja et al. 2017; Margutti et al. 2017; Haggard et al. 2017; Ruan et al. 2018; Pooley et al. 2017) and a radio signal (Hallinan et al. 2017; Alexander et al. 2017; Mooley et al. 2018) were discovered at the position of the optical transient after $\sim 9$ days and $\sim 16$ days, respectively. A slow multi-wavelength flux-rise of the non-thermal emission was observed until $\sim 150$ days (Lyman et al. 2018; Margutti et al. 2018; Troja et al. 2018) before entering a flattening-decaying phase (D'Avanzo et al. 2018; Dobie et al. 2018; Alexander et al. 2018; Hajela et al. 2019; Fong et al. 2019). Very Long Baseline Interferometry observations enabled measurement of the superluminal proper motion of the radio counterpart (Mooley et al. 2018) and constrained the apparent size of the source (Ghirlanda et al. 2019), proving that a relativistic and narrowly-collimated jet successfully emerged from the neutron star merger. These multimessenger observations support the hypothesis that GW170817 came from a BNS coalescence, which was the source of the short GRB 170817A (Goldstein et al. 2017a; Savchenko et al. 2017a) and of the kilonova powered by the radioactive decay of r-process nuclei produced in the collision (Pian et al. 2017; McCully et al. 2017; Smartt et al. 2017; Chornock et al. 2017; Nicholl et al. 2017; Shappee et al. 2017; Kasliwal et al. 2017; Evans et al. 2017).

## 4 Public alerts

To facilitate the rapid identification of electromagnetic or neutrino counterparts to GW detections, and to maximize the science that the entire scientific community can do with them, GW candidate events are released as public alerts as of the start of O3.[17]

Within minutes of detection *Preliminary GCN Notices* are issued automatically for a candidate that satisfies pre-established criteria. After each *Preliminary GCN Notice*, a Rapid Response Team (RRT), composed of staff from the detector sites, the analysis teams, the detector characterization team, and the low-latency follow-up team, are called upon to confirm or retract the candidate on the basis of semi-automated detector characterization and data quality checks. Events which are expected to be electromagnetically bright such as BNS or NSBH mergers require vetting by the full RRT. BBH mergers are also inspected by the RRT but the issuance of a circular or retraction may have a latency of up to 1 day. For non-BBH events our goal is to issue an *Initial GCN Notice* accompanied by either a *GCN Circular*, or a *Retraction GCN Notice* within a few hours.

[17] Documentation is available in the LIGO/Virgo Public Alerts User Guide at emfollow.docs.ligo.org/userguide/.

Interesting events, which do not satisfy our criteria for issuing an automatic alert are discussed in *ad hoc* daily meetings. Alerts generated by such events may have a latency on the order of 1 day.

*Update GCN Notices* and *Circulars* are issued whenever further analysis leads to improved estimates of the source localization, significance, or classification. Localization updates are sent until the position is determined more accurately by public announcement of an unambiguous counterpart. Figure 8 shows the timeline of the different types of *GCN Notices* after a GW signal. *Update GCN Notices* and *Circulars* may be issued hours, days, or even weeks after the event.

### 4.1 O3 false alarm rate threshold for automatic alerts

The FAR threshold to release automatic alerts for CBC events targets an overall astrophysical purity of 90% across all categories of mergers. Different classes of CBCs may individually have higher or lower purity than 90%. This 90% purity translates to a FAR threshold of 1/(2 months) for CBC. For the unmodeled burst events the FAR threshold is 1/year. Single detector CBC candidates, which are found in coincidence with a multi-messenger source, must still satisfy the FAR threshold of 1/(2 months) in order to generate an automatic alert. In general multiple pipelines search for CBC and Burst candidates. Individual FAR thresholds for each pipeline are corrected by a trials factor, so that the overall FAR thresholds described above are satisfied for each class of event.

### 4.2 Alert contents

The alert contains information to support the search for counterparts including:

- A candidate identifier, which can be used to examine the event properties in the Gravitational Wave Candidate Event Database.[18]
- The FAR of the candidate in Hz.
- The localization given as a posterior probability distribution of the source's sky position. For CBC events, we send a 3-D sky map, which also contains the direction-dependent luminosity distance. The localization is encoded as a HEALPIX projection in FITS file format.
- For Burst candidates the central frequency in Hz, the duration in seconds and the GW fluence in erg/cm$^2$.
- For CBC candidates the probability $p_{\text{astro}}$, that the signal is astrophyiscal (see Sect. 3.1). This probability comes from evaluating whether the source belongs to one of five categories: BNS merger (both component masses $< 3\,M_{\odot}$), MassGap merger ($3\,M_{\odot} <$ one component mass $< 5\,M_{\odot}$) NSBH merger (one component mass $< 3\,M_{\odot}$ and the other $> 5\,M_{\odot}$), BBH merger (both component masses $> 5\,M_{\odot}$), Terrestrial (i.e. Noise). Details about the formalism used to compute

[18] gracedb.ligo.org.

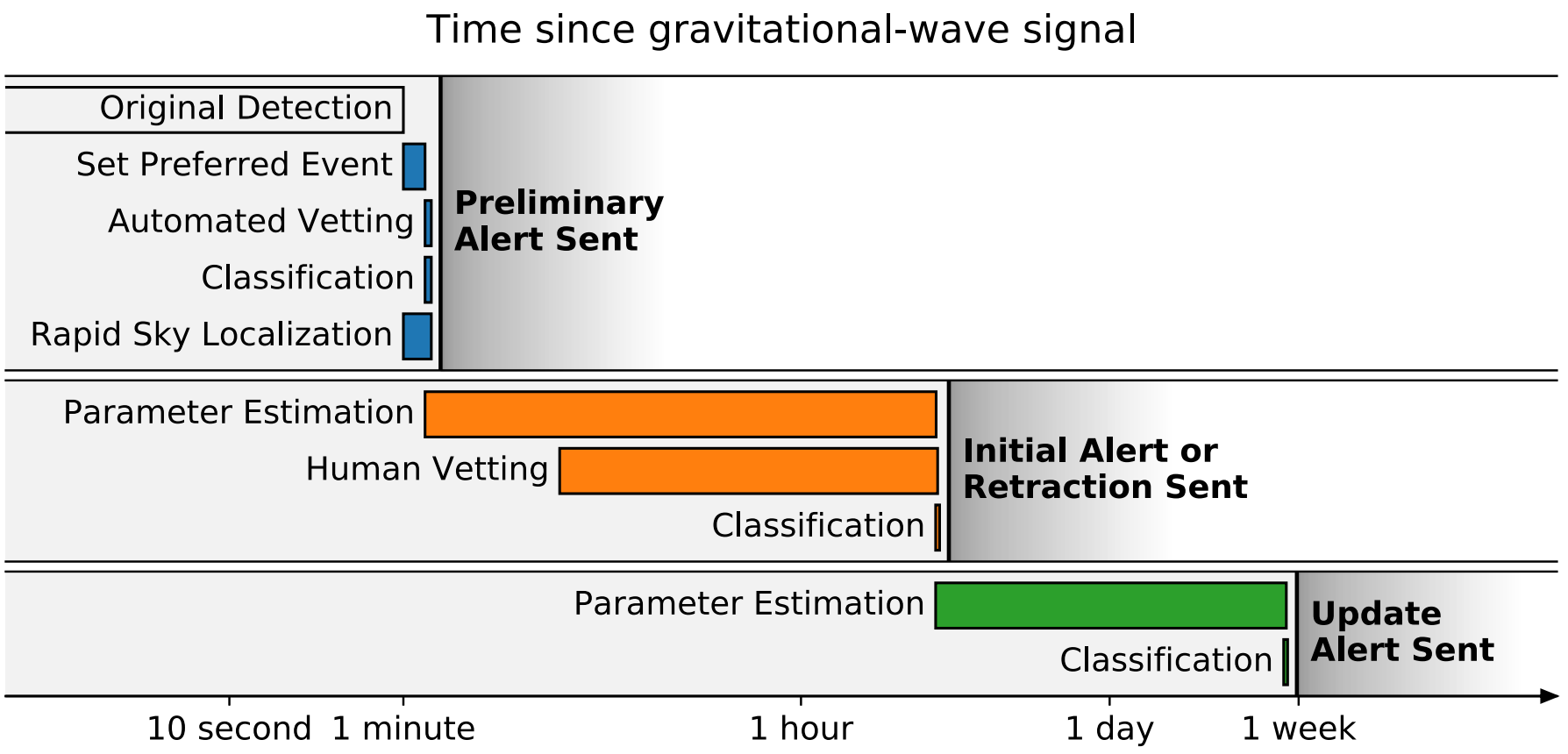

**Fig. 8** Alert timeline. The *Preliminary GCN Notice* is sent autonomously within 1–10 min after the GW candidate trigger time. Some preliminary alerts may be retracted after human inspection for data quality, instrumental conditions, and pipeline behavior. The human vetted *Initial GCN Notice* or *Retraction GCN Notice* and associated *GCN Circular* are distributed within a few hours for BNS or NSBH sources and within 1 day for BBH. Update notices and circulars are sent whenever the estimate of the parameters of the signal significantly improves. Image adapted from the LIGO/Virgo Public Alerts User Guide (see footnote 17)

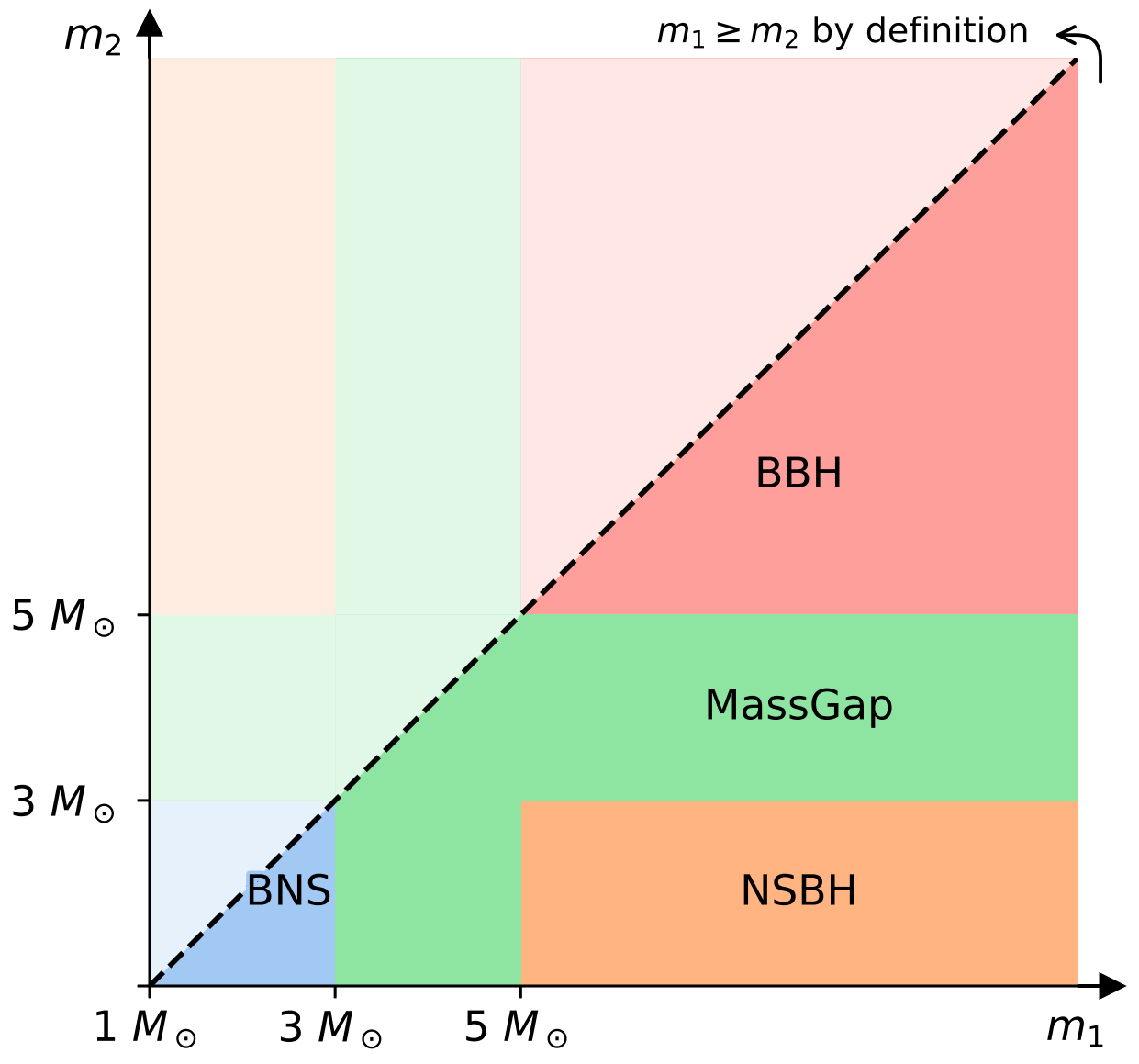

**Fig. 9** The four astrophysical categories in terms (BNS, NSBH, BBH, and MassGap) of component masses *m1* and *m2*, which are used to define the source classification. By convention, the component masses are defined such that $m1 \geqslant m2$, so that the primary compact object in the binary (i.e., component 1), is always more massive than the secondary compact object (i.e., component 2). Image adapted from the LIGO/Virgo Public Alerts User Guide (see footnote 17)

**Table 5** Expected BNS, BBH and NSBH detections and localization accuracy for the O3 and O4 observing runs

| Observation run | Network | Expected BNS detections | Expected NSBH detections | Expected BBH detections |
|---|---|---|---|---|
| O3 | HLV | $1^{+12}_{-1}$ | $0^{+19}_{-0}$ | $17^{+22}_{-11}$ |
| O4 | HLVK | $10^{+52}_{-10}$ | $1^{+91}_{-1}$ | $79^{+89}_{-44}$ |
| | | Area ($\mathrm{deg}^2$) 90% c.r. | Area ($\mathrm{deg}^2$) 90% c.r. | Area ($\mathrm{deg}^2$) 90% c.r. |
| O3 | HLV | $270^{+34}_{-20}$ | $330^{+24}_{-31}$ | $280^{+30}_{-23}$ |
| O4 | HLVK | $33^{+5}_{-5}$ | $50^{+8}_{-8}$ | $41^{+7}_{-6}$ |
| | | Comoving volume ($10^3$ $\mathrm{Mpc}^3$) 90% c.r. | Comoving volume ($10^3$ $\mathrm{Mpc}^3$) 90% c.r. | Comoving volume ($10^3$ $\mathrm{Mpc}^3$) 90% c.r. |
| O3 | HLV | $120^{+19}_{-24}$ | $860^{+150}_{-150}$ | $16000^{+2200}_{-2500}$ |
| O4 | HLVK | $52^{+10}_{-9}$ | $430^{+100}_{-78}$ | $7700^{+1500}_{-920}$ |

Results are shown for the three-detector HLV network in O3 and the four-detector HLVK network in O4. The detection number predictions are given as detection counts in a one-calendar-year observing run; the quoted confidence intervals combine the log-normal uncertainty in the merger rate with Poisson counting statistics. The localization accuracy is given as the median 90% credible area and median 90% credible comoving volume; their confidence intervals describe Monte Carlo uncertainty from the simulation. All quantities are given as 90% credible intervals of the form $x^{+b}_{-a}$, where $x$ is the 50th percentile, $(x-a)$ is the 5th percentile, and $(x+b)$ is the 95th percentile this probability are given in Kapadia et al. (2020). The method to assign probabilities of astrophysical origin to GW candidate events is based on redistributing, via mass-based template weighting, the foreground probabilities of candidate events with respect to the background model across the astrophysical categories shown in Fig. 9. The template weights are computed from injection campaigns of astrophysical sources with defined mass and spin distributions into the detector data, and recovering them via a detection pipeline. The method accuracy depends on how well the template weights are constructed. Kapadia et al. (2020) show that the constructed weights were adequate and the method works well for the GW signals observed during O1 and O2. Using template weights that are not perfectly constructed for the O3 signals would not affect distinguishing astrophysical vs terrestrial probability, but could make the method imprecise in distinguishing among low-mass systems containing one or two neutron stars, or two low mass black holes.

- For CBC candidates the probability that one or both components has a mass consistent with a neutron star (HasNS), that is a mass $<3\,M_{\odot}$. And the probability that the system ejected a non-zero amount of neutron star matter (HasRemnant). This latter evaluates the probability that baryon mass is left outside the merger remnant using the masses and spins of the binary system inferred from the signal (Foucart 2012; Pannarale and Ohme 2014; Foucart et al. 2018).

*GCN Circulars* and *Updates* may also include a concise description of any instrument or data quality issues that could affect the significance estimate, the localization, and the GW parameter inferences.

### 4.3 O3a gravitational-wave candidate alerts

The first half of the third observing run of Advanced LIGO and Virgo, O3a, began at 1500 UTC on April 1, 2019 and lasted 6 months. During O3a, 41 gravitational-wave candidate events were publicly released in low-latency; 8 were retracted, 3 have a larger probability to be classified as terrestrial, 3 as BNS, 2 as lying in the mass gap, 4 as NSBH systems, and 21 as BBH systems. Among the GW candidates classified as astrophysical, 19 have a FAR smaller than 1/10 years. The median sky-localization 90% credible area for BBH (the systems for which we have larger statistics) is around 400 $\mathrm{deg}^2$. However, the sky localization and numbers of O3a GW candidates cannot be directly compared to the predictions of the present paper due to the more conservative SNR threshold used in our simulation to define a detection. The smaller SNR threshold for releasing alerts is expected to give larger sky localization and higher detection counts with respect to the ones quoted in Table 5. Larger sky localization in O3 is also expected due to the release of signals detected during the observations of a single interferometer (while our simulation requires a detection of SNR $> 4$ in at least two instruments). Two single-detector GW candidates (classified as BNS and NSBH) were released in low-latency with a sky localization covering several thousands of $\mathrm{deg}^2$. For 21 alerts an updated LALInference sky map was sent (within a week for 17 of them and with a larger latency for the others). The LALInference localizations (90% c.r.) resulted to be in median smaller by about 40% with respect to the initial BAYESTAR localizations. For a few cases, while the BAYESTAR localization was bimodal and weights the two modes equally, LALInference favors one localization over the other. These differences among BAYESTAR and LALInference are attributable to a multiplicative factor introduced into the on-line BAYESTAR pipeline to account for estimation errors from search pipelines (see Singer and Price 2016). Removing the multiplicative factor results in agreement between the O3a BAYESTAR and LALInference localizations.

## 5 Observing scenarios

In this section we present an estimate of the expected number of BNS, NSBH and BBH detections for the three-detector HLV network in O3 and for the four-detector HLVK network in O4. We also summarize the expected localization area and comoving volume obtained with the simulation described in Sect. 3.2.1. The expectations for the number of events we will detect in each source category comes from the same simulation of populations used to evaluate the localization capability.

The astrophysical parameter distribution, detector duty cycle, and detection threshold are described in Sect. 3.2.1.

In contrast to previous versions of this paper where we gave the range of estimated rates per unit time, here, we evaluate the plausible detection counts per one-calendar-year observing run. We model each source category as a Poisson process combined with the source rate densities and anticipated surveyed volume, and we marginalize over the uncertainty in the source rate estimates. This procedure allows us to incorporate the counting uncertainty from the Poisson process, but makes forming an exact 90% confidence interval impossible, and as such, these intervals overcover. All source categories assume parameterized physical property distributions[19] for which the chosen parameters (e.g., power laws or mass limits) are consistent with current measurements and their uncertainties (Abbott et al. 2018a). We assume constant rate density in comoving volume and source-frame time. For BNS we use the source rate density 110–3840 $\mathrm{Gpc}^{-3}\,\mathrm{year}^{-1}$ from Abbott et al. (2018c) and Abbott et al. (2018a).[20] For BBH we use the rate calculated using Model B in Abbott et al. (2018a), 25–109 $\mathrm{Gpc}^{-3}\,\mathrm{year}^{-1}$, and for NSBH we use the rate from Abadie et al. (2010b), 0.6–1000 $\mathrm{Gpc}^{-3}\,\mathrm{year}^{-1}$.[21] There are numerous uncertainties involved in the component mass and spin distributions for NSBH systems and this is reflected in our estimates for expected detections. The rate is obtained assuming that NSBH mergers exist, but the absence of this type of system cannot be excluded by the O1 and O2 GW observations.

As described in in Sect. 3.2.1, we assume a duty factor of 70% for each detector, uncorrelated between instruments, and we require a network SNR of at least 12 and an SNR $> 4$ in at least two instruments.[22] All SNRs are calculated assuming perfect templates. Event significance is established not solely by SNR, but by ranking statistics used by the detection pipelines which also use the goodness of fit and the rate of background in the ranking (Cannon et al. 2015; Usman et al. 2016; Nitz et al. 2017). The thresholds set on the ranking statistic propagate to the inferred search volume $VT$, where $V$ is the spacetime volumes surveyed per unit detector time defined in Sect. 2, and $T$ is the observing time incorporating the effects of the detectors duty cycles. Our estimates are realistic projections, but the search volume is sensitive to our assumptions on source population, detection criteria and network characteristics. The simulation results for the HLV network in O3 and the HLVK network in O4 are summarized in Table 5. Adding KAGRA to the network in O3 does not change the detection counts. The results are given for a population of sources with aligned and anti-aligned spins; there is no significant change of the detection counts using isotropic spin distributions. Using uniform mass distributions

[19] Details on the adopted distributions of the source properties are given in Sect. 3.2.1.

[20] This rate combines rate intervals estimated with uniform mass and Gaussian mass distribution populations (See Sect. 3). While this does not represent a physical distribution of sources, it does incorporate a degree of uncertainty arising from our ignorance of the actual BNS distribution.

[21] We do not limit the rate to the O1–O2 upper limit of 610 $\mathrm{Gpc}^{-3}\,\mathrm{year}^{-1}$ obtained with point mass assumptions in order to consider a broad distribution of masses.

[22] This is a conservative choice since we routinely detect events with lower SNR (see Abbott et al. 2018c).

(instead of a Gaussian distribution for NS and a power-law distribution for BH) increases the counts in Table 5 by about 30% for BNSs and 60% NSBHs.

### 5.1 O3: aLIGO 110–130 Mpc, AdV 50 Mpc, KAGRA 8–25 Mpc

This year long run began in April 2019 with the three detector HLV network and with KAGRA planning to join in the latter stages. The simulation to estimate the number of expected GW detections uses the curves in Fig. 1 for the two aLIGO and the AdV detectors, corresponding to a BNS range of 130 Mpc, 110 Mpc, and 50 Mpc respectively. For KAGRA we use the 25 Mpc curve.

The BNS search volume $VT$ is evaluated to be $3.3 \times 10^6$ Mpc$^3$ year with $1^{+12}_{-1}$ expected detections. The median 90% credible region for the localization area (volume) of BNS is $270^{+34}_{-20}$ deg$^2$ ($120^{+19}_{-24} \times 10^3$ Mpc$^3$).[23] A percentage of 9–13% (2–4%) of the events are expected to have a 90% credible region smaller than 20 deg$^2$ (5 deg$^2$). For BBH the search volume $VT$ is $3.4 \times 10^8$ Mpc$^3$ year, and the expected detections are $17^{+22}_{-11}$. The median 90% credible region for the localization area (volume) is $280^{+30}_{-23}$ deg$^2$ ($16000^{+2200}_{-2500} \times 10^3$ Mpc$^3$). A percentage of 9–13% (2–3%) of the events are expected to have a 90% credible area smaller than 20 deg$^2$ (5 deg$^2$).

### 5.2 O4: aLIGO 160–190 Mpc, AdV 90–120 Mpc, KAGRA 25–130 Mpc

O4 is planned to have a duration of 1 year. The aLIGO detectors will be near their design sensitivity, with a BNS range of 160–190 Mpc. AdV will have completed Phase 1 of the AdV+ upgrade with an anticipated BNS range of 90–120 Mpc. As the newest member of the network, KAGRA has the largest uncertainty in projected O4 sensitivity, a BNS range of 25–130 Mpc. For estimating the number of events expected to be detected in O4 we use an intermediate sensitivity curve for KAGRA, one with a BNS range of 80 Mpc, and the target sensitivity curve (the highest O4 sensitivity) for aLIGO and for AdV.

In O4 we predict a BNS search volume $VT$ of $1.6 \times 10^7$ Mpc$^3$ year, and $10^{+52}_{-10}$ expected detections. The median 90% credible region for the localization area (volume) of BNS is $33^{+5}_{-5}$ deg$^2$ ($52^{+10}_{-9} \times 10^3$ Mpc$^3$). A percentage of 38–44% (12–16%) of the events are expected to have a 90% credible region smaller than 20 deg$^2$ (5 deg$^2$). For BBH the VT searched is 1.5 Gpc$^3$ year with $79^{+89}_{-44}$ expected detections. The median 90% credible region for the localization area (volume) of BBH is $41^{+7}_{-6}$ deg$^2$ ($7700^{+1500}_{-920} \times 10^3$ Mpc$^3$). A percentage of 35–39% (11–14%) of the events are expected to have a 90% credible area smaller than 20 deg$^2$ (5 deg$^2$).

Table 5 lists the results described above for O3 and O4, including also predictions for NSBH. Localization capabilities of unmodelled searches for BBHs and IMBHB are shown in Sect. 3.2.2, where we give also the BBH and IMBHB ranges for the unmodeled search algorithm cWB in Table 4.

[23] The median area and volume are given as 90% Monte Carlo sampling confidence bounds on the median 90% credible regions.

### 5.3 O5: aLIGO (LIGO-India will join in 2025) 330 Mpc, AdV 150–260 Mpc, KAGRA 130+ Mpc

There is considerable uncertainty in looking this far ahead. The current plan envisions the aLIGO instruments, including an instrument in India in 2025, beginning observations after the A+ upgrade (Abbott et al. 2018d), the AdV instrument participating after the completion of the AdV+ upgrade (Phase 2), and KAGRA operating at or above its final O4 sensitivity of 130+ Mpc. In Fig. 2 we show target sensitivities for this phase of observations. In practice the detectors are likely to begin observations at a lower sensitivity and then gradually improve over the span of several years. For now we make no quantitative predictions about the expected performance of the GW network in this era.

For O3, O4 and O5, Table 2 gives the ranges for BNS, NSBH, and BBH, and for generic burst sources emitting $10^{-2}\,M_\odot c^2$ and $10^{-9}\,M_\odot c^2$ in GWs.

## 6 Conclusions

We have presented our current best estimate of the plausible observing scenarios for the network of Advanced GW detectors, including aLIGO, AdV, and KAGRA. This includes plans, already approved and in progress, to upgrade the aLIGO and AdV instruments. We outlined the observing schedule and sensitivity evolution for the next decade, showing the anticipated strain sensitivities and the corresponding range at which we can detect BNSs, BBHs, NSBHs, and unmodeled signals. We evaluated our ability to localize BNSs, BBHs, NSBHs, and IMBHBs using matched-filter and unmodelled searches. For BNSs, BBHs, and NSBHs systems we estimated the number of expected detections in a one-calendar-year observing run. We detailed our plan to automatically notify the astronomical community of event candidates, starting in O3. This information will help to optimize multi-messenger follow-up and source identification, to plan instrument operation and projects, and to evaluate joint detections in order to maximize the science return of each GW detection (e.g., Abadie et al. 2012b; Aasi et al. 2014a; Kasliwal and Nissanke 2014; Singer et al. 2014; Cannon et al. 2012; Evans et al. 2016a; Gehrels et al. 2016; Ghosh et al. 2016; Chan et al. 2017; Rana et al. 2017; Patricelli et al. 2016; Salafia et al. 2017; Patricelli et al. 2018; Coughlin et al. 2018; Vinciguerra et al. 2019).

The three-detector aLIGO and AdV network has demonstrated the ability to localize signals to sky areas of a few tens of square degrees. The addition of KAGRA, and later LIGO-India to the network will improve this situation further. While the median sky localization area is expected to be a few hundreds of square degrees for all types of binary systems in O3, it will improve to be a few tens of square degrees during O4. By 2025 a five-detector network consisting of three upgraded LIGO detectors in the United States and India, an upgraded Virgo detector, and possibly an upgraded KAGRA instrument is expected to operate at sensitivities approaching twice that of their predecessors, and a median sky localization area of a few degrees. Detection of BBHs will become routine. A few hundred BBH detections will allow us to probe the major formation channel, and distinguish between isolated binaries and systems formed in star clusters (see e.g, Zevin et al. 2017; Stevenson et al. 2017; Farr et al. 2017). BNSs are expected to be detected with a rate from a few per year, to a few per month. Associated electromagnetic counterparts will probe properties of relativistic jets and sub-relativistic dynamical ejecta, the nucleosynthesis of heavy elements, and will enable precise cosmology.

The scenarios described here are our best current projections, they will evolve as detector installation and commissioning progress. Regular updates are planned to ensure that the content remains timely and relevant.

**Acknowledgements** The authors gratefully acknowledge the support of the United States National Science Foundation (NSF) for the construction and operation of the LIGO Laboratory and Advanced LIGO as well as the Science and Technology Facilities Council (STFC) of the United Kingdom, the Max Planck Society (MPS), and the State of Niedersachsen/Germany for support of the construction of Advanced LIGO and construction and operation of the GEO600 detector. Additional support for Advanced LIGO was provided by the Australian Research Council. The authors gratefully acknowledge the Italian Istituto Nazionale di Fisica Nucleare (INFN), the French Centre National de la Recherche Scientifique (CNRS) and the Foundation for Fundamental Research on Matter supported by the Netherlands Organisation for Scientific Research, for the construction and operation of the Virgo detector and the creation and support of the EGO consortium. The authors also gratefully acknowledge research support from these agencies as well as by the Council of Scientific and Industrial Research of India, the Department of Science and Technology, India, the Science and Engineering Research Board (SERB), India, the Ministry of Human Resource Development, India, the Spanish Agencia Estatal de Investigación, the Vicepresidència i Conselleria d'Innovació, Recerca i Turisme and the Conselleria d'Educació i Universitat del Govern de les Illes Balears, the Conselleria d'Educació, Investigació, Cultura i Esport de la Generalitat Valenciana, the National Science Centre of Poland, the Swiss National Science Foundation (SNSF), the Russian Foundation for Basic Research, the Russian Science Foundation, the European Commission, the European Regional Development Funds (ERDF), the Royal Society, the Scottish Funding Council, the Scottish Universities Physics Alliance, the Hungarian Scientific Research Fund (OTKA), the Lyon Institute of Origins (LIO), the Paris Île-de-France Region, the National Research, Development and Innovation Office Hungary (NKFIH), the National Research Foundation of Korea, Industry Canada and the Province of Ontario through the Ministry of Economic Development and Innovation, the Natural Science and Engineering Research Council Canada, the Canadian Institute for Advanced Research, the Brazilian Ministry of Science, Technology, Innovations, and Communications, the International Center for Theoretical Physics South American Institute for Fundamental Research (ICTP-SAIFR), the Research Grants Council of Hong Kong, the National Natural Science Foundation of China (NSFC), the Leverhulme Trust, the Research Corporation, the Ministry of Science and Technology (MOST), Taiwan and the Kavli Foundation. The authors gratefully acknowledge the support of the NSF, STFC, INFN and CNRS for provision of computational resources. This work was supported by MEXT, JSPS Leading-edge Research Infrastructure Program, JSPS Grant-in-Aid for Specially Promoted Research 26000005, JSPS Grant-in-Aid for Scientific Research on Innovative Areas 2905: JP17H06358, JP17H06361 and JP17H06364, JSPS Core-to-Core Program A. Advanced Research Networks, JSPS Grant-in-Aid for Scientific Research (S) 17H06133, the joint research program of the Institute for Cosmic Ray Research, University of Tokyo in Japan, National Research Foundation (NRF) and Computing Infrastructure Project of KISTI-GSDC in Korea, Academia Sinica (AS), AS Grid Center (ASGC) and the Ministry of Science and Technology (MoST) in Taiwan under grants including AS-CDA-105-M06. This article has been assigned LIGO Document number P1200087, Virgo Document number VIR-0288A-12, and KAGRA Document number JGW-P1808427.

## A Changes between versions

Since publication of the previous version (Abbott et al. 2018f), several updates to the document have been made. The most significant changes are that we now frame our projections in terms of observing runs, we include final results from O2, and we updated our localization projections to include KAGRA as a fourth detector. Key differences are outlined below.

### A.1 Updates to Sect. 2, "Construction, commissioning and observing phases":

1. The observing roadmap is now discussed in terms of observing runs rather than the "Early","Mid", "Late" nomenclature used in previous versions.
2. The O1 and O2 discussion happens earlier in the section. Future planned runs are discussed at the end. Discussion of O1 and O2 duty cycle now occurs in this section.
3. A subsection has been added for O3.
4. Table 2 and Fig. 1 have been updated to include the actual performance in O1 and O2, and in the first months of O3 (which started April 1st 2019 and is ongoing). The projected performance is given for O4 and O5.
5. Table 2 also includes ranges for NSBH and Burst sources.
6. There is now a discussion, with projected sensitivities, of upgrades to aLIGO and AdV.
7. Figure 2 now extends past 2026, showing LIGO-India joining the network.

### A.2 Updates to Sect. 3, "Searches for gravitational-wave transients":

1. We include the latest O2 results from (Abbott et al. 2018c, a).
2. The discussion is considerably shortened compared to the previous version.
3. There is a new subsection describing the O1 and O2 follow-up program.
4. New localization simulations have been performed for three-detector and four-detector networks at O3 and O4 sensitivities. Results are presented for both CBC and Burst signals.
5. The CBC simulation used astrophysically motivated populations of sources with properties consistent with the O1 and O2 results.
6. CBC signal sky-localization now includes luminosity distance and comoving volume in addition to area.
7. The anticipated sky-localization is given as before for BNS systems and additionally for NSBH and BBH systems.
8. The burst simulation used astrophysically motivated populations of BBHs and IMBHBs in contrast to the previous version which included a variety of generic waveform morphologies.

9. Figure 3 has new results from O2 and updated results from O1.
10. Figure 4 is updated to show the effect of adding KAGRA to the network.
11. Figure 5 shows sky maps of the confident GW events detected during O1 and O2 (Abbott et al. 2018c, 2019d) by the low-latency and full offline analysis. The previous version of this figure showed the sky location for a simulated BNS signal.
12. Table 3 is new. It shows luminosity distance and localization of the O1 and O2 confident detections obtained by the low-latency and full offline analysis.
13. Figure 6 has updated localization plots for compact binary mergers (BNS, BBH, NSBH) in O3 and O4. This includes also luminosity distance and comoving volume expectations. The figure no longer shows the performance of LALInference, which is evaluated to be consistent with BAYESTAR. The previous version of this figure had results for BNS systems alone.
14. Figure 7 has updated localization plots for burst sources in O3 and O4. The figure no longer shows the umodeled search performance for generic waveform morphologies, but for BBH and IMBHB signals.
15. Table 4 is new; it shows the range of the cWB searches for BBH and IMBHB mergers.

### A. 3 Section 4, "Public alerts"

1. This is a new section describing how alerts are issued publicly and automatically starting starting from O3.
2. Includes a brief discussion of alerts sent during O3a.

### A. 4 Updates to Sect. 5, "Observing scenarios":

1. The scenarios are now discussed in terms of observing runs up to O5.
2. Discussion of the O1 and O2 runs has been moved to earlier in the paper.
3. New simulations have been performed for the expected number of detections in O3 and O4. We give the range of plausible detection counts in a one-calendar-year observing run instead of range of estimated rates per unit time as given in the previous version. The detection expectations are given also for NSBH and BBH mergers.
4. The rate simulation uses source properties and astrophysical rates consistent with the O1 and O2 results.
5. Table 5, which replaces Table 3 in the previous version, has been updated significantly. In particular, we no longer quote ranges since these are reported in Sect. 2. We show anticipated numbers for O3 and O4 only; prior run information is no longer reported here. We added estimates of comoving volume localization, and information for BBH and NSBH mergers.

## Affiliations

**B. P. Abbott[1] · R. Abbott[1] · T. D. Abbott[2] · S. Abraham[3] · F. Acernese[4,5] · K. Ackley[6] · C. Adams[7] · V. B. Adya[8,9] · C. Affeldt[8,9] · M. Agathos[10] · K. Agatsuma[11] · N. Aggarwal[12] · O. D. Aguiar[13] · L. Aiello[14,15] · A. Ain[3] · P. Ajith[16] · T. Akutsu[17] · G. Allen[18] · A. Allocca[19,20] · M. A. Aloy[21] · P. A. Altin[22] · A. Amato[23] · A. Ananyeva[1] · S. B. Anderson[1] · W. G. Anderson[24] · M. Ando[25,26] · S. V. Angelova[27] · S. Antier[28] · S. Appert[1] · K. Arai[1] · Koya Arai[29] · Y. Arai[29] · S. Araki[30] · A. Araya[31] · M. C. Araya[1] · J. S. Areeda[32] · M. Arène[33] · N. Aritomi[25] · N. Arnaud[28,34] · K. G. Arun[35] · S. Ascenzi[36,37] · G. Ashton[6] · Y. Aso[38,39] · S. M. Aston[7] · P. Astone[40] · F. Aubin[41] · P. Aufmuth[9] · K. AultONeal[42] · C. Austin[2] · V. Avendano[43] · A. Avila-Alvarez[32] · S. Babak[33,44] · P. Bacon[33] · F. Badaracco[14,15] · M. K. M. Bader[45] · S. W. Bae[46] · Y. B. Bae[47] · L. Baiotti[48] · R. Bajpai[49] · P. T. Baker[50] · F. Baldaccini[51,52] · G. Ballardin[34] · S. W. Ballmer[53] · S. Banagiri[54] · J. C. Barayoga[1] · S. E. Barclay[55] · B. C. Barish[1] · D. Barker[56] · K. Barkett[57] · S. Barnum[12] · F. Barone[4,5] ·**

B. Barr[55] · L. Barsotti[12] · M. Barsuglia[33] · D. Barta[58] · J. Bartlett[56] ·
M. A. Barton[17,55] · I. Bartos[59] · R. Bassiri[60] · A. Basti[19,20] · M. Bawaj[52,61] ·
J. C. Bayley[55] · M. Bazzan[62,63] · B. Bécsy[64] · M. Bejger[33,65] · I. Belahcene[28] ·
A. S. Bell[55] · D. Beniwal[66] · B. K. Berger[60] · G. Bergmann[8,9] · S. Bernuzzi[67,68] ·
J. J. Bero[69] · C. P. L. Berry[70] · D. Bersanetti[71] · A. Bertolini[45] ·
J. Betzwieser[7] · R. Bhandare[72] · J. Bidler[32] · I. A. Bilenko[73] · S. A. Bilgili[50] ·
G. Billingsley[1] · J. Birch[7] · R. Birney[27] · O. Birnholtz[69] · S. Biscans[1,12] ·
S. Biscoveanu[6] · A. Bisht[9] · M. Bitossi[20,34] · M. A. Bizouard[28] · J. K. Blackburn[1] ·
C. D. Blair[7] · D. G. Blair[74] · R. M. Blair[56] · S. Bloemen[75] · N. Bode[8,9] ·
M. Boer[76] · Y. Boetzel[77] · G. Bogaert[76] · F. Bondu[78] · E. Bonilla[60] ·
R. Bonnand[41] · P. Booker[8,9] · B. A. Boom[45] · C. D. Booth[79] · R. Bork[1] ·
V. Boschi[34] · S. Bose[3,80] · K. Bossie[7] · V. Bossilkov[74] · J. Bosveld[74] ·
Y. Bouffanais[33] · A. Bozzi[34] · C. Bradaschia[20] · P. R. Brady[24] · A. Bramley[7] ·
M. Branchesi[14,15] · J. E. Brau[81] · T. Briant[82] · J. H. Briggs[55] · F. Brighenti[83,84] ·
A. Brillet[76] · M. Brinkmann[8,9] · V. Brisson[28] · P. Brockill[24] · A. F. Brooks[1] ·
D. A. Brown[53] · D. D. Brown[66] · S. Brunett[1] · A. Buikema[12] · T. Bulik[85] ·
H. J. Bulten[45,86] · A. Buonanno[44,87] · D. Buskulic[41] · C. Buy[33] ·
R. L. Byer[60] · M. Cabero[8,9] · L. Cadonati[88] · G. Cagnoli[23,89] · C. Cahillane[1] ·
J. Calderón Bustillo[6] · T. A. Callister[1] · E. Calloni[5,90] · J. B. Camp[91] ·
W. A. Campbell[6] · M. Canepa[71,92] · K. Cannon[26] · K. C. Cannon[26] ·
H. Cao[66] · J. Cao[93] · E. Capocasa[17,33] · F. Carbognani[34] · S. Caride[94] ·
M. F. Carney[70] · G. Carullo[19] · J. Casanueva Diaz[20] · C. Casentini[36,37] ·
S. Caudill[45] · M. Cavaglià[95] · F. Cavalier[28] · R. Cavalieri[34] · G. Cella[20] ·
P. Cerdá-Durán[21] · G. Cerretani[19,20] · E. Cesarini[37,96] · O. Chaibi[76] ·
K. Chakravarti[3] · S. J. Chamberlin[97] · M. Chan[55] · M. L. Chan[98] ·
S. Chao[99] · P. Charlton[100] · E. A. Chase[70] · E. Chassande-Mottin[33] ·
D. Chatterjee[24] · M. Chaturvedi[72] · K. Chatziioannou[101] · B. D. Cheeseboro[50] ·
C. S. Chen[102] · H. Y. Chen[103] · K. H. Chen[104] · X. Chen[74] · Y. Chen[57] ·
Y. R. Chen[105,106] · H.-P. Cheng[59] · C. K. Cheong[107] · H. Y. Chia[59] ·
A. Chincarini[71] · A. Chiummo[34] · G. Cho[108] · H. S. Cho[109] · M. Cho[87] ·
N. Christensen[76,110] · H. Y. Chu[104] · Q. Chu[74] · Y. K. Chu[111] · S. Chua[82] ·
K. W. Chung[107] · S. Chung[74] · G. Ciani[62,63] · A. A. Ciobanu[66] · R. Ciolfi[112,113] ·
F. Cipriano[76] · A. Cirone[71,92] · F. Clara[56] · J. A. Clark[88] · P. Clearwater[114] ·
F. Cleva[76] · C. Cocchieri[95] · E. Coccia[14,15] · P.-F. Cohadon[82] · D. Cohen[28] ·
R. Colgan[115] · M. Colleoni[116] · C. G. Collette[117] · C. Collins[11] ·
L. R. Cominsky[118] · M. Constancio Jr.[13] · L. Conti[63] · S. J. Cooper[11] ·
P. Corban[7] · T. R. Corbitt[2] · I. Cordero-Carrión[119] · K. R. Corley[115] ·
N. Cornish[64] · A. Corsi[94] · S. Cortese[34] · C. A. Costa[13] · R. Cotesta[44] ·
M. W. Coughlin[1] · S. B. Coughlin[70,79] · J.-P. Coulon[76] · S. T. Countryman[115] ·
P. Couvares[1] · P. B. Covas[116] · E. E. Cowan[88] · D. M. Coward[74] ·
M. J. Cowart[7] · D. C. Coyne[1] · R. Coyne[120] · J. D. E. Creighton[24] ·
T. D. Creighton[121] · J. Cripe[2] · M. Croquette[82] · S. G. Crowder[122] ·
T. J. Cullen[2] · A. Cumming[55] · L. Cunningham[55] · E. Cuoco[34] ·
T. Dal Canton[91] · G. Dálya[123] · S. L. Danilishin[8,9] · S. D'Antonio[37] ·
K. Danzmann[8,9] · A. Dasgupta[124] · C. F. Da Silva Costa[59] · L. E. H. Datrier[55] ·
V. Dattilo[34] · I. Dave[72] · M. Davier[28] · D. Davis[53] · E. J. Daw[125] ·

D. DeBra[60] · M. Deenadayalan[3] · J. Degallaix[23] · M. De Laurentis[5,90] ·
S. Deléglise[82] · W. Del Pozzo[19,20] · L. M. DeMarchi[70] · N. Demos[12] ·
T. Dent[8,9,126] · R. De Pietri[68,127] · J. Derby[32] · R. De Rosa[5,90] ·
C. De Rossi[23,34] · R. DeSalvo[128] · O. de Varona[8,9] · S. Dhurandhar[3] ·
M. C. Díaz[121] · T. Dietrich[45] · L. Di Fiore[5] · M. Di Giovanni[113,129] ·
T. Di Girolamo[5,90] · A. Di Lieto[19,20] · B. Ding[117] · S. Di Pace[40,130] ·
I. Di Palma[40,130] · F. Di Renzo[19,20] · A. Dmitriev[11] · Z. Doctor[103] ·
K. Doi[131] · F. Donovan[12] · K. L. Dooley[79,95] · S. Doravari[8,9] · I. Dorrington[79] ·
T. P. Downes[24] · M. Drago[14,15] · J. C. Driggers[56] · Z. Du[93] · J.-G. Ducoin[28] ·
P. Dupej[55] · S. E. Dwyer[56] · P. J. Easter[6] · T. B. Edo[125] · M. C. Edwards[110] ·
A. Effler[7] · S. Eguchi[98] · P. Ehrens[1] · J. Eichholz[1] · S. S. Eikenberry[59] ·
M. Eisenmann[41] · R. A. Eisenstein[12] · Y. Enomoto[25] · R. C. Essick[103] ·
H. Estelles[116] · D. Estevez[41] · Z. B. Etienne[50] · T. Etzel[1] · M. Evans[12] ·
T. M. Evans[7] · V. Fafone[14,36,37] · H. Fair[53] · S. Fairhurst[79] · X. Fan[93] ·
S. Farinon[71] · B. Farr[81] · W. M. Farr[11] · E. J. Fauchon-Jones[79] ·
M. Favata[43] · M. Fays[125] · M. Fazio[132] · C. Fee[133] · J. Feicht[1] ·
M. M. Fejer[60] · F. Feng[33] · A. Fernandez-Galiana[12] · I. Ferrante[19,20] ·
E. C. Ferreira[13] · T. A. Ferreira[13] · F. Ferrini[34] · F. Fidecaro[19,20] ·
I. Fiori[34] · D. Fiorucci[33] · M. Fishbach[103] · R. P. Fisher[53,134] · J. M. Fishner[12] ·
M. Fitz-Axen[54] · R. Flaminio[17,41] · M. Fletcher[55] · E. Flynn[32] · H. Fong[101] ·
J. A. Font[21,135] · P. W. F. Forsyth[22] · J.-D. Fournier[76] · S. Frasca[40,130] ·
F. Frasconi[20] · Z. Frei[123] · A. Freise[11] · R. Frey[81] · V. Frey[28] · P. Fritschel[12] ·
V. V. Frolov[7] · Y. Fujii[136] · M. Fukunaga[29] · M. Fukushima[137] ·
P. Fulda[59] · M. Fyffe[7] · H. A. Gabbard[55] · B. U. Gadre[3] · S. M. Gaebel[11] ·
J. R. Gair[138] · L. Gammaitoni[51] · M. R. Ganija[66] · S. G. Gaonkar[3] ·
A. Garcia[32] · C. García-Quirós[116] · F. Garufi[5,90] · B. Gateley[56] ·
S. Gaudio[42] · G. Gaur[139] · V. Gayathri[140] · G. G. Ge[141] · G. Gemme[71] ·
E. Genin[34] · A. Gennai[20] · D. George[18] · J. George[72] · L. Gergely[142] ·
V. Germain[41] · S. Ghonge[88] · Abhirup Ghosh[16] · Archisman Ghosh[45] ·
S. Ghosh[24] · B. Giacomazzo[113,129] · J. A. Giaime[2,7] · K. D. Giardina[7] ·
A. Giazotto[20] · K. Gill[42] · G. Giordano[4,5] · L. Glover[128] · P. Godwin[97] ·
E. Goetz[56] · R. Goetz[59] · B. Goncharov[6] · G. González[2] ·
J. M. Gonzalez Castro[19,20] · A. Gopakumar[143] · M. L. Gorodetsky[73] ·
S. E. Gossan[1] · M. Gosselin[34] · R. Gouaty[41] · A. Grado[5,144] · C. Graef[55] ·
M. Granata[23] · A. Grant[55] · S. Gras[12] · P. Grassia[1] · C. Gray[56] ·
R. Gray[55] · G. Greco[83,84] · A. C. Green[11,59] · R. Green[79] · E. M. Gretarsson[42] ·
P. Groot[75] · H. Grote[79] · S. Grunewald[44] · P. Gruning[28] · G. M. Guidi[83,84] ·
H. K. Gulati[124] · Y. Guo[45] · A. Gupta[97] · M. K. Gupta[124] · E. K. Gustafson[1] ·
R. Gustafson[145] · L. Haegel[116] · A. Hagiwara[146] · S. Haino[147] ·
O. Halim[14,15] · B. R. Hall[80] · E. D. Hall[12] · E. Z. Hamilton[79] · G. Hammond[55] ·
M. Haney[77] · M. M. Hanke[8,9] · J. Hanks[56] · C. Hanna[97] · M. D. Hannam[79] ·
O. A. Hannuksela[107] · J. Hanson[7] · T. Hardwick[2] · K. Haris[16] · J. Harms[14,15] ·
G. M. Harry[148] · I. W. Harry[44] · K. Hasegawa[29] · C.-J. Haster[101] ·
K. Haughian[55] · H. Hayakawa[149] · K. Hayama[98] · F. J. Hayes[55] ·
J. Healy[69] · A. Heidmann[82] · M. C. Heintze[7] · H. Heitmann[76] ·
P. Hello[28] · G. Hemming[34] · M. Hendry[55] · I. S. Heng[55] · J. Hennig[8,9] ·

A. W. Heptonstall[1] · M. Heurs[8,9] · S. Hild[55] · Y. Himemoto[150] ·
T. Hinderer[45,151,152] · Y. Hiranuma[153] · N. Hirata[17] · E. Hirose[29] ·
D. Hoak[34] · S. Hochheim[8,9] · D. Hofman[23] · A. M. Holgado[18] · N. A. Holland[22] ·
K. Holt[7] · D. E. Holz[103] · Z. Hong[111] · P. Hopkins[79] · C. Horst[24] ·
J. Hough[55] · E. J. Howell[74] · C. G. Hoy[79] · A. Hreibi[76] · B. H. Hsieh[154] ·
G. Z. Huang[111] · P. W. Huang[141] · Y. J. Huang[147] · E. A. Huerta[18] ·
D. Huet[28] · B. Hughey[42] · M. Hulko[1] · S. Husa[116] · S. H. Huttner[55] ·
T. Huynh-Dinh[7] · B. Idzkowski[85] · A. Iess[36,37] · B. Ikenoue[137] ·
S. Imam[111] · K. Inayoshi[155] · C. Ingram[66] · Y. Inoue[104] · R. Inta[94] ·
G. Intini[40,130] · K. Ioka[156] · B. Irwin[133] · H. N. Isa[55] · J.-M. Isac[82] ·
M. Isi[1] · Y. Itoh[157,158] · B. R. Iyer[16] · K. Izumi[56,159] · T. Jacqmin[82] ·
S. J. Jadhav[160] · K. Jani[88] · N. N. Janthalur[160] · P. Jaranowski[161] ·
A. C. Jenkins[162] · J. Jiang[59] · D. S. Johnson[18] · A. W. Jones[11] · D. I. Jones[163] ·
R. Jones[55] · R. J. G. Jonker[45] · L. Ju[74] · K. Jung[164] · P. Jung[149] ·
J. Junker[8,9] · T. Kajita[29] · C. V. Kalaghatgi[79] · V. Kalogera[70] · B. Kamai[1] ·
M. Kamiizumi[149] · N. Kanda[157,158] · S. Kandhasamy[95] · G. W. Kang[46] ·
J. B. Kanner[1] · S. J. Kapadia[24] · S. Karki[81] · K. S. Karvinen[8,9] · R. Kashyap[16] ·
M. Kasprzack[1] · S. Katsanevas[34] · E. Katsavounidis[12] · W. Katzman[7] ·
S. Kaufer[9] · K. Kawabe[56] · K. Kawaguchi[29] · N. Kawai[165] · T. Kawasaki[25] ·
N. V. Keerthana[3] · F. Kéfélian[76] · D. Keitel[55] · R. Kennedy[125] ·
J. S. Key[166] · F. Y. Khalili[73] · H. Khan[32] · I. Khan[14,37] · S. Khan[8,9] ·
Z. Khan[124] · E. A. Khazanov[167] · M. Khursheed[72] · N. Kijbunchoo[22] ·
Chunglee Kim[168] · C. Kim[169] · J. C. Kim[170] · J. Kim[171] · K. Kim[107] ·
W. Kim[66] · W. S. Kim[47] · Y.-M. Kim[164] · C. Kimball[70] · N. Kimura[146] ·
E. J. King[66] · P. J. King[56] · M. Kinley-Hanlon[148] · R. Kirchhoff[8,9] ·
J. S. Kissel[56] · N. Kita[25] · H. Kitazawa[131] · L. Kleybolte[172] · J. H. Klika[24] ·
S. Klimenko[59] · T. D. Knowles[50] · E. Knyazev[12] · P. Koch[8,9] ·
S. M. Koehlenbeck[8,9] · G. Koekoek[45,173] · Y. Kojima[174] · K. Kokeyama[149] ·
S. Koley[45] · K. Komori[25] · V. Kondrashov[1] · A. K. H. Kong[105,106] ·
A. Kontos[12] · N. Koper[8,9] · M. Korobko[172] · W. Z. Korth[1] · K. Kotake[98] ·
I. Kowalska[85] · D. B. Kozak[1] · C. Kozakai[146] · R. Kozu[175] · V. Kringel[8,9] ·
N. Krishnendu[35] · A. Królak[176,177] · G. Kuehn[8,9] · A. Kumar[160] ·
P. Kumar[178] · Rahul Kumar[1] · R. Kumar[124] · S. Kumar[16] · J. Kume[26] ·
C. M. Kuo[104] · H. S. Kuo[111] · L. Kuo[99] · S. Kuroyanagi[179] · K. Kusayanagi[165] ·
A. Kutynia[176] · K. Kwak[164] · S. Kwang[24] · B. D. Lackey[44] · K. H. Lai[107] ·
T. L. Lam[107] · M. Landry[56] · B. B. Lane[12] · R. N. Lang[180] · J. Lange[69] ·
B. Lantz[60] · R. K. Lanza[12] · A. Lartaux-Vollard[28] · P. D. Lasky[6] ·
M. Laxen[7] · A. Lazzarini[1] · C. Lazzaro[63] · P. Leaci[40,130] · S. Leavey[8,9] ·
Y. K. Lecoeuche[56] · C. H. Lee[109] · H. K. Lee[181] · H. M. Lee[182] · H. W. Lee[170] ·
J. Lee[108] · K. Lee[55] · R. K. Lee[105,106] · J. Lehmann[8,9] · A. Lenon[50] ·
M. Leonardi[17] · N. Leroy[28] · N. Letendre[41] · Y. Levin[6,115] · J. Li[93] ·
K. J. L. Li[107] · T. G. F. Li[107] · X. Li[57] · C. Y. Lin[183] · F. Lin[6] · F. L. Lin[111] ·
L. C. C. Lin[164] · F. Linde[45] · S. D. Linker[128] · T. B. Littenberg[184] ·
G. C. Liu[102] · J. Liu[74] · X. Liu[24] · R. K. L. Lo[1,107] · N. A. Lockerbie[27] ·
L. T. London[79] · A. Longo[185,186] · M. Lorenzini[14,15] · V. Loriette[187] ·
M. Lormand[7] · G. Losurdo[20] · J. D. Lough[8,9] · C. O. Lousto[69] · G. Lovelace[32] ·

M. E. Lower[188] · H. Lück[8,9] · D. Lumaca[36,37] · A. P. Lundgren[189] ·
L. W. Luo[147] · R. Lynch[12] · Y. Ma[57] · R. Macas[79] · S. Macfoy[27] ·
M. MacInnis[12] · D. M. Macleod[79] · A. Macquet[76] · F. Magaña-Sandoval[53] ·
L. Magaña Zertuche[95] · R. M. Magee[97] · E. Majorana[40] · I. Maksimovic[187] ·
A. Malik[72] · N. Man[76] · V. Mandic[54] · V. Mangano[55] · G. L. Mansell[12,56] ·
M. Manske[22,24] · M. Mantovani[34] · F. Marchesoni[52,61] · M. Marchio[17] ·
F. Marion[41] · S. Márka[115] · Z. Márka[115] · C. Markakis[10,18] · A. S. Markosyan[60] ·
A. Markowitz[1] · E. Maros[1] · A. Marquina[119] · S. Marsat[44] · F. Martelli[83,84] ·
I. W. Martin[55] · R. M. Martin[43] · D. V. Martynov[11] · K. Mason[12] ·
E. Massera[125] · A. Masserot[41] · T. J. Massinger[1] · M. Masso-Reid[55] ·
S. Mastrogiovanni[40,130] · A. Matas[44,54] · F. Matichard[1,12] · L. Matone[115] ·
N. Mavalvala[12] · N. Mazumder[80] · J. J. McCann[74] · R. McCarthy[56] ·
D. E. McClelland[22] · S. McCormick[7] · L. McCuller[12] · S. C. McGuire[190] ·
J. McIver[1] · D. J. McManus[22] · T. McRae[22] · S. T. McWilliams[50] ·
D. Meacher[97] · G. D. Meadors[6] · M. Mehmet[8,9] · A. K. Mehta[16] ·
J. Meidam[45] · A. Melatos[114] · G. Mendell[56] · R. A. Mercer[24] · L. Mereni[23] ·
E. L. Merilh[56] · M. Merzougui[76] · S. Meshkov[1] · C. Messenger[55] ·
C. Messick[97] · R. Metzdorff[82] · P. M. Meyers[114] · H. Miao[11] · C. Michel[23] ·
Y. Michimura[25,26] · H. Middleton[114] · E. E. Mikhailov[191] · L. Milano[5,90] ·
A. L. Miller[59] · A. Miller[40,130] · M. Millhouse[64] · J. C. Mills[79] ·
M. C. Milovich-Goff[128] · O. Minazzoli[76,192] · Y. Minenkov[37] · N. Mio[193] ·
A. Mishkin[59] · C. Mishra[194] · T. Mistry[125] · S. Mitra[3] · V. P. Mitrofanov[73] ·
G. Mitselmakher[59] · R. Mittleman[12] · O. Miyakawa[149] · A. Miyamoto[157] ·
Y. Miyazaki[25] · K. Miyo[149] · S. Miyoki[149] · G. Mo[110] · D. Moffa[133] ·
K. Mogushi[95] · S. R. P. Mohapatra[12] · M. Montani[83,84] · C. J. Moore[10] ·
D. Moraru[56] · G. Moreno[56] · S. Morisaki[26] · Y. Moriwaki[131] · B. Mours[41] ·
C. M. Mow-Lowry[11] · Arunava Mukherjee[8,9] · D. Mukherjee[24] ·
S. Mukherjee[121] · N. Mukund[3] · A. Mullavey[7] · J. Munch[66] · E. A. Muñiz[53] ·
M. Muratore[42] · P. G. Murray[55] · K. Nagano[29] · S. Nagano[195] ·
A. Nagar[96,196,197] · K. Nakamura[17] · H. Nakano[198] · M. Nakano[29] ·
R. Nakashima[165] · I. Nardecchia[36,37] · T. Narikawa[199] · L. Naticchioni[40,130] ·
R. K. Nayak[200] · R. Negishi[153] · J. Neilson[128] · G. Nelemans[45,75] ·
T. J. N. Nelson[7] · M. Nery[8,9] · A. Neunzert[145] · K. Y. Ng[12] · S. Ng[66] ·
P. Nguyen[81] · W. T. Ni[105,141] · D. Nichols[45,151] · A. Nishizawa[26,201] ·
S. Nissanke[45,151] · F. Nocera[34] · C. North[79] · L. K. Nuttall[189] · M. Obergaulinger[21] ·
J. Oberling[56] · B. D. O'Brien[59] · Y. Obuchi[137] · G. D. O'Dea[128] ·
W. Ogaki[29] · G. H. Ogin[202] · J. J. Oh[47] · S. H. Oh[47] · M. Ohashi[149] ·
N. Ohishi[38] · M. Ohkawa[203] · F. Ohme[8,9] · H. Ohta[26] · M. A. Okada[13] ·
K. Okutomi[39] · M. Oliver[116] · K. Oohara[153] · C. P. Ooi[25] · P. Oppermann[8,9] ·
Richard J. Oram[7] · B. O'Reilly[7] · R. G. Ormiston[54] · L. F. Ortega[59] ·
R. O'Shaughnessy[69] · S. Oshino[149] · S. Ossokine[44] · D. J. Ottaway[66] ·
H. Overmier[7] · B. J. Owen[94] · A. E. Pace[97] · G. Pagano[19,20] · M. A. Page[74] ·
A. Pai[140] · S. A. Pai[72] · J. R. Palamos[81] · O. Palashov[167] · C. Palomba[40] ·
A. Pal-Singh[172] · Huang-Wei Pan[99] · K. C. Pan[105,106] · B. Pang[57] ·
H. F. Pang[104] · P. T. H. Pang[107] · C. Pankow[70] · F. Pannarale[40,130] ·
B. C. Pant[72] · F. Paoletti[20] · A. Paoli[34] · M. A. Papa[8,24,44] · A. Parida[3] ·

J. Park[204] · W. Parker[7,190] · D. Pascucci[55] · A. Pasqualetti[34] · R. Passaquieti[19,20] · D. Passuello[20] · M. Patil[177] · B. Patricelli[19,20] · B. L. Pearlstone[55] · C. Pedersen[79] · M. Pedraza[1] · R. Pedurand[23,205] · A. Pele[7] · F. E. Peña Arellano[149] · S. Penn[206] · C. J. Perez[56] · A. Perreca[113,129] · H. P. Pfeiffer[44,101] · M. Phelps[8,9] · K. S. Phukon[3] · O. J. Piccinni[40,130] · M. Pichot[76] · F. Piergiovanni[83,84] · G. Pillant[34] · L. Pinard[23] · I. Pinto[207] · M. Pirello[56] · M. Pitkin[55] · R. Poggiani[19,20] · D. Y. T. Pong[107] · S. Ponrathnam[3] · P. Popolizio[34] · E. K. Porter[33] · J. Powell[188] · A. K. Prajapati[124] · J. Prasad[3] · K. Prasai[60] · R. Prasanna[160] · G. Pratten[116] · T. Prestegard[24] · S. Privitera[44] · G. A. Prodi[113,129] · L. G. Prokhorov[73] · O. Puncken[8,9] · M. Punturo[52] · P. Puppo[40] · M. Pürrer[44] · H. Qi[24] · V. Quetschke[121] · P. J. Quinonez[42] · E. A. Quintero[1] · R. Quitzow-James[81] · F. J. Raab[56] · H. Radkins[56] · N. Radulescu[76] · P. Raffai[123] · S. Raja[72] · C. Rajan[72] · B. Rajbhandari[94] · M. Rakhmanov[121] · K. E. Ramirez[121] · A. Ramos-Buades[116] · Javed Rana[3] · K. Rao[70] · P. Rapagnani[40,130] · V. Raymond[79] · M. Razzano[19,20] · J. Read[32] · T. Regimbau[41] · L. Rei[71] · S. Reid[27] · D. H. Reitze[1,59] · W. Ren[18] · F. Ricci[40,130] · C. J. Richardson[42] · J. W. Richardson[1] · P. M. Ricker[18] · K. Riles[145] · M. Rizzo[70] · N. A. Robertson[1,55] · R. Robie[55] · F. Robinet[28] · A. Rocchi[37] · L. Rolland[41] · J. G. Rollins[1] · V. J. Roma[81] · M. Romanelli[78] · R. Romano[4,5] · C. L. Romel[56] · J. H. Romie[7] · K. Rose[133] · D. Rosińska[65,208] · S. G. Rosofsky[18] · M. P. Ross[209] · S. Rowan[55] · A. Rüdiger[8,9] · P. Ruggi[34] · G. Rutins[210] · K. Ryan[56] · S. Sachdev[1] · T. Sadecki[56] · N. Sago[211] · S. Saito[137] · Y. Saito[149] · K. Sakai[212] · Y. Sakai[153] · H. Sakamoto[131] · M. Sakellariadou[162] · Y. Sakuno[98] · L. Salconi[34] · M. Saleem[35] · A. Samajdar[45] · L. Sammut[6] · E. J. Sanchez[1] · L. E. Sanchez[1] · N. Sanchis-Gual[21] · V. Sandberg[56] · J. R. Sanders[53] · K. A. Santiago[43] · N. Sarin[6] · B. Sassolas[23] · B. S. Sathyaprakash[79,97] · S. Sato[213] · T. Sato[203] · O. Sauter[145] · R. L. Savage[56] · T. Sawada[157] · P. Schale[81] · M. Scheel[57] · J. Scheuer[70] · P. Schmidt[75] · R. Schnabel[172] · R. M. S. Schofield[81] · A. Schönbeck[172] · E. Schreiber[8,9] · B. W. Schulte[8,9] · B. F. Schutz[79] · S. G. Schwalbe[42] · J. Scott[55] · S. M. Scott[22] · E. Seidel[18] · T. Sekiguchi[26] · Y. Sekiguchi[214] · D. Sellers[7] · A. S. Sengupta[215] · N. Sennett[44] · D. Sentenac[34] · V. Sequino[14,36,37] · A. Sergeev[167] · Y. Setyawati[8,9] · D. A. Shaddock[22] · T. Shaffer[56] · M. S. Shahriar[70] · M. B. Shaner[128] · L. Shao[44] · P. Sharma[72] · P. Shawhan[87] · H. Shen[18] · S. Shibagaki[98] · R. Shimizu[137] · T. Shimoda[25] · K. Shimode[149] · R. Shink[216] · H. Shinkai[217] · T. Shishido[49] · A. Shoda[17] · D. H. Shoemaker[12] · D. M. Shoemaker[88] · S. ShyamSundar[72] · K. Siellez[88] · M. Sieniawska[65] · D. Sigg[56] · A. D. Silva[13] · L. P. Singer[91] · N. Singh[85] · A. Singhal[14,40] · A. M. Sintes[116] · S. Sitmukhambetov[121] · V. Skliris[79] · B. J. J. Slagmolen[22] · T. J. Slaven-Blair[74] · J. R. Smith[32] · R. J. E. Smith[6] · S. Somala[218] · K. Somiya[165] · E. J. Son[47] · B. Sorazu[55] · F. Sorrentino[71] · H. Sotani[219] · T. Souradeep[3] · E. Sowell[94] · A. P. Spencer[55] · A. K. Srivastava[124] · V. Srivastava[53] · K. Staats[70] · C. Stachie[76] · M. Standke[8,9] · D. A. Steer[33] · M. Steinke[8,9] · J. Steinlechner[55,172] · S. Steinlechner[172] · D. Steinmeyer[8,9] · S. P. Stevenson[188] · D. Stocks[60] · R. Stone[121] · D. J. Stops[11] · K. A. Strain[55] ·

G. Stratta[83,84] · S. E. Strigin[73] · A. Strunk[56] · R. Sturani[220] · A. L. Stuver[221] · V. Sudhir[12] · R. Sugimoto[131] · T. Z. Summerscales[222] · L. Sun[1] · S. Sunil[124] · J. Suresh[3] · P. J. Sutton[79] · Takamasa Suzuki[203] · Toshikazu Suzuki[29] · B. L. Swinkels[45] · M. J. Szczepańczyk[42] · M. Tacca[45] · H. Tagoshi[29] · S. C. Tait[55] · H. Takahashi[223] · R. Takahashi[17] · A. Takamori[31] · S. Takano[25] · H. Takeda[25] · M. Takeda[153] · C. Talbot[6] · D. Talukder[81] · H. Tanaka[154] · Kazuyuki Tanaka[157] · Kenta Tanaka[154] · Taiki Tanaka[29] · Takahiro Tanaka[199] · S. Tanioka[17,39] · D. B. Tanner[59] · M. Tápai[142] · E. N. Tapia San Martin[17] · A. Taracchini[44] · J. D. Tasson[110] · R. Taylor[1] · S. Telada[224] · F. Thies[8,9] · M. Thomas[7] · P. Thomas[56] · S. R. Thondapu[72] · K. A. Thorne[7] · E. Thrane[6] · Shubhanshu Tiwari[113,129] · Srishti Tiwari[143] · V. Tiwari[79] · K. Toland[55] · T. Tomaru[146] · Y. Tomigami[157] · T. Tomura[149] · M. Tonelli[19,20] · Z. Tornasi[55] · A. Torres-Forné[225] · C. I. Torrie[1] · D. Töyrä[11] · F. Travasso[34,52,61] · G. Traylor[7] · M. C. Tringali[85] · A. Trovato[33] · L. Trozzo[20,149,226] · R. Trudeau[1] · K. W. Tsang[45] · T. T. L. Tsang[107] · M. Tse[12] · R. Tso[57] · K. Tsubono[25] · S. Tsuchida[157] · L. Tsukada[26] · D. Tsuna[26] · T. Tsuzuki[137] · D. Tuyenbayev[121,147] · N. Uchikata[227] · T. Uchiyama[149] · A. Ueda[146] · T. Uehara[59,228] · K. Ueno[26] · G. Ueshima[223] · D. Ugolini[229] · C. S. Unnikrishnan[143] · F. Uraguchi[137] · A. L. Urban[2] · T. Ushiba[29] · S. A. Usman[79] · H. Vahlbruch[9] · G. Vajente[1] · G. Valdes[2] · N. van Bakel[45] · M. van Beuzekom[45] · J. F. J. van den Brand[45,86] · C. Van Den Broeck[45,230] · D. C. Vander-Hyde[53] · L. van der Schaaf[45] · J. V. van Heijningen[74] · M. H. P. M. van Putten[231] · A. A. van Veggel[55] · M. Vardaro[62,63] · V. Varma[57] · S. Vass[1] · M. Vasúth[58] · A. Vecchio[11] · G. Vedovato[63] · J. Veitch[55] · P. J. Veitch[66] · K. Venkateswara[209] · G. Venugopalan[1] · D. Verkindt[41] · F. Vetrano[83,84] · A. Viceré[83,84] · A. D. Viets[24] · D. J. Vine[210] · J.-Y. Vinet[76] · S. Vitale[12] · Francisco Hernandez Vivanco[6] · T. Vo[53] · H. Vocca[51,52] · C. Vorvick[56] · S. P. Vyatchanin[73] · A. R. Wade[1] · L. E. Wade[133] · M. Wade[133] · R. Walet[45] · M. Walker[32] · L. Wallace[1] · S. Walsh[24] · G. Wang[14,20] · H. Wang[11] · J. Wang[141] · J. Z. Wang[145] · W. H. Wang[121] · Y. F. Wang[107] · R. L. Ward[22] · Z. A. Warden[42] · J. Warner[56] · M. Was[41] · J. Watchi[117] · B. Weaver[56] · L.-W. Wei[8,9] · M. Weinert[8,9] · A. J. Weinstein[1] · R. Weiss[12] · F. Wellmann[8,9] · L. Wen[74] · E. K. Wessel[18] · P. Weßels[8,9] · J. W. Westhouse[42] · K. Wette[22] · J. T. Whelan[69] · B. F. Whiting[59] · C. Whittle[12] · D. M. Wilken[8,9] · D. Williams[55] · A. R. Williamson[45,151] · J. L. Willis[1] · B. Willke[8,9] · M. H. Wimmer[8,9] · W. Winkler[8,9] · C. C. Wipf[1] · H. Wittel[8,9] · G. Woan[55] · J. Woehler[8,9] · J. K. Wofford[69] · J. Worden[56] · J. L. Wright[55] · C. M. Wu[105,106] · D. S. Wu[8,9] · H. C. Wu[105,106] · S. R. Wu[105,106] · D. M. Wysocki[69] · L. Xiao[1] · W. R. Xu[111] · T. Yamada[154] · H. Yamamoto[1] · Kazuhiro Yamamoto[131] · Kohei Yamamoto[154] · T. Yamamoto[149] · C. C. Yancey[87] · L. Yang[132] · M. J. Yap[22] · M. Yazback[59] · D. W. Yeeles[79] · K. Yokogawa[131] · J. Yokoyama[25,26] · T. Yokozawa[149] · T. Yoshioka[131] · Hang Yu[12] · Haocun Yu[12] · S. H. R. Yuen[107] · H. Yuzurihara[29] · M. Yvert[41] · A. K. Zadrożny[121,176] · M. Zanolin[42] · S. Zeidler[17] · T. Zelenova[34] · J.-P. Zendri[63] · M. Zevin[70] · J. Zhang[74] · L. Zhang[1] · T. Zhang[55] ·

**C. Zhao[74] · Y. Zhao[17] · M. Zhou[70] · Z. Zhou[70] · X. J. Zhu[6] · Z. H. Zhu[232] · A. B. Zimmerman[233] · M. E. Zucker[1,12] · J. Zweizig[1]**

V. Brisson: Deceased, February 2018.

A. Giazotto: Deceased, November 2017.

A. Rüdiger: Deceased, July 2018.

[1] LIGO, California Institute of Technology, Pasadena, CA 91125, USA

[2] Louisiana State University, Baton Rouge, LA 70803, USA

[3] Inter-University Centre for Astronomy and Astrophysics, Pune 411007, India

[4] Università di Salerno, 84084 Fisciano, Salerno, Italy

[5] INFN, Sezione di Napoli, Complesso Universitario di Monte S.Angelo, 80126 Napoli, Italy

[6] OzGrav, School of Physics and Astronomy, Monash University, Clayton, VIC 3800, Australia

[7] LIGO Livingston Observatory, Livingston, LA 70754, USA

[8] Max Planck Institute for Gravitational Physics (Albert Einstein Institute), 30167 Hannover, Germany

[9] Leibniz Universität Hannover, 30167 Hannover, Germany

[10] University of Cambridge, Cambridge CB2 1TN, UK

[11] University of Birmingham, Birmingham B15 2TT, UK

[12] LIGO, Massachusetts Institute of Technology, Cambridge, MA 02139, USA

[13] Instituto Nacional de Pesquisas Espaciais, 12227-010 São José dos Campos, São Paulo, Brazil

[14] Gran Sasso Science Institute (GSSI), 67100 L'Aquila, Italy

[15] INFN, Laboratori Nazionali del Gran Sasso, 67100 Assergi, Italy

[16] International Centre for Theoretical Sciences, Tata Institute of Fundamental Research, Bengaluru 560089, India

[17] National Astronomical Observatory of Japan (NAOJ), 2-21-1,Osawa, Mitaka-shi, Tokyo 181-8588, Japan

[18] NCSA, University of Illinois at Urbana-Champaign, Urbana, IL 61801, USA

[19] Università di Pisa, 56127 Pisa, Italy

[20] INFN, Sezione di Pisa, 56127 Pisa, Italy

[21] Departamento de Astronomía y Astrofísica, Universitat de València, 46100 Burjassot, València, Spain

[22] OzGrav, Australian National University, Canberra, ACT 0200, Australia

[23] Laboratoire des Matériaux Avancés (LMA), CNRS/IN2P3, 69622 Villeurbanne, France

[24] University of Wisconsin-Milwaukee, Milwaukee, WI 53201, USA

[25] Department of Physics, The University of Tokyo, 7-3-1, Hongo, Bunkyo-ku, Tokyo 113-0033, Japan

[26] Research Center for the Early Universe (RESCEU), The University of Tokyo, 7-3-1, Hongo, Bunkyo-ku, Tokyo 113-0033, Japan

[27] SUPA, University of Strathclyde, Glasgow, G1 1XQ, UK

[28] LAL, Univ. Paris-Sud, CNRS/IN2P3, Université Paris-Saclay, 91898 Orsay, France

[29] Institute for Cosmic Ray Research (ICRR), The University of Tokyo, 5-1-5 Kashiwa-no-Ha, Kashiwa City, Chiba 277-8582, Japan

[30] Accelerator Laboratory, High Energy Accelerator Research Organization (KEK), 1-1, Oho, Tsukuba-shi, Ibaraki 305-0801, Japan

[31] Earthquake Research Institute, The University of Tokyo, 1-1-1,Yayoi, Bunkyo-ku, Tokyo 113-0032, Japan

[32] California State University Fullerton, Fullerton, CA 92831, USA

[33] APC, AstroParticule et Cosmologie, Université Paris Diderot, CNRS/IN2P3, CEA/Irfu, Observatoire de Paris, Sorbonne Paris Cité, 75205 Paris Cedex 13, France

[34] European Gravitational Observatory (EGO), 56021 Cascina, Pisa, Italy

[35] Chennai Mathematical Institute, Chennai 603103, India

[36] Università di Roma Tor Vergata, 00133 Rome, Italy

[37] INFN, Sezione di Roma Tor Vergata, 00133 Rome, Italy

[38] National Astronomical Observatory of Japan (NAOJ), Kamioka Branch, 238 Higashi-Mozumi, Kamioka-cho, Hida-shi, Gifu Pref., Japan

[39] The Graduate University for Advanced Studies (SOKENDAI), 2-21-1, Osawa, Mitaka-shi, Tokyo 181-8588, Japan

[40] INFN, Sezione di Roma, 00185 Rome, Italy

[41] Laboratoire d'Annecy de Physique des Particules (LAPP), CNRS/IN2P3, Univ. Grenoble Alpes, Université Savoie Mont Blanc, 74941 Annecy, France

[42] Embry-Riddle Aeronautical University, Prescott, AZ 86301, USA

[43] Montclair State University, Montclair, NJ 07043, USA

[44] Max Planck Institute for Gravitational Physics (Albert Einstein Institute), 14476 Potsdam-Golm, Germany

[45] Nikhef, Science Park 105, 1098 XG Amsterdam, The Netherlands

[46] Korea Institute of Science and Technology Information (KISTI), 245 Daehak-ro, Yuseong-gu, Daejeon 34141, Korea

[47] National Institute for Mathematical Sciences, 70, Yuseong-daero 1689 Beon-gil, Yuseong-gu, Daejeon 34047, Korea

[48] Graduate School of Science, Osaka University, 1-1, Machikaneyama-cho, Toyonaka-shi, Osaka 560-0043, Japan

[49] School of High Energy Accelerator Science, The Graduate University for Advanced Studies (SOKENDAI), 1-1 Oho, Tsukuba-shi, Ibaraki 305-0801, Japan

[50] West Virginia University, Morgantown, WV 26506, USA

[51] Università di Perugia, 06123 Perugia, Italy

[52] INFN, Sezione di Perugia, 06123 Perugia, Italy

[53] Syracuse University, Syracuse, NY 13244, USA

[54] University of Minnesota, Minneapolis, MN 55455, USA

[55] SUPA, University of Glasgow, Glasgow G12 8QQ, UK

[56] LIGO Hanford Observatory, Richland, WA 99352, USA

[57] Caltech CaRT, Pasadena, CA 91125, USA

[58] Wigner RCP, RMKI, Konkoly Thege Miklós út 29-33, Budapest 1121, Hungary

[59] University of Florida, Gainesville, FL 32611, USA

[60] Stanford University, Stanford, CA 94305, USA

[61] Università di Camerino, Dipartimento di Fisica, 62032 Camerino, Italy

[62] Dipartimento di Fisica e Astronomia, Università di Padova, 35131 Padova, Italy

[63] INFN, Sezione di Padova, 35131 Padova, Italy

[64] Montana State University, Bozeman, MT 59717, USA

[65] Nicolaus Copernicus Astronomical Center, Polish Academy of Sciences, 00-716 Warsaw, Poland

[66] OzGrav, University of Adelaide, Adelaide, SA 5005, Australia

[67] Theoretisch-Physikalisches Institut, Friedrich-Schiller-Universität Jena, 07743 Jena, Germany

[68] INFN, Sezione di Milano Bicocca, Gruppo Collegato di Parma, 43124 Parma, Italy

[69] Rochester Institute of Technology, Rochester, NY 14623, USA

[70] Center for Interdisciplinary Exploration and Research in Astrophysics (CIERA), Northwestern University, Evanston, IL 60208, USA

[71] INFN, Sezione di Genova, 16146 Genova, Italy

[72] RRCAT, Indore, Madhya Pradesh 452013, India

[73] Faculty of Physics, Lomonosov Moscow State University, Moscow, Russia 119991

[74] OzGrav, University of Western Australia, Crawley, WA 6009, Australia

[75] Department of Astrophysics/IMAPP, Radboud University Nijmegen, P.O. Box 9010, 6500 GL Nijmegen, The Netherlands

[76] Artemis, Observatoire Côte d’Azur, CNRS, Université Côte d’Azur, CS 34229, 06304 Nice Cedex 4, France

[77] Physik-Institut, University of Zurich, Winterthurerstrasse 190, 8057 Zurich, Switzerland

[78] Univ Rennes, CNRS, Institut FOTON - UMR6082, 3500 Rennes, France

[79] Cardiff University, Cardiff CF24 3AA, UK

[80] Washington State University, Pullman, WA 99164, USA

[81] University of Oregon, Eugene, OR 97403, USA

[82] Laboratoire Kastler Brossel, Sorbonne Université, CNRS,ENS-Université PSL, Collège de France, 75005 Paris, France

[83] Università degli Studi di Urbino ’Carlo Bo’, 61029 Urbino, Italy

[84] INFN, Sezione di Firenze, 50019 Sesto Fiorentino, Firenze, Italy

[85] Astronomical Observatory Warsaw University, 00-478 Warsaw, Poland

[86] VU University Amsterdam, 1081 HV Amsterdam, The Netherlands

[87] University of Maryland, College Park, MD 20742, USA

[88] School of Physics, Georgia Institute of Technology, Atlanta, GA 30332, USA

[89] Université Claude Bernard Lyon 1, 69622 Villeurbanne, France

[90] Università di Napoli ’Federico II’, Complesso Universitario di Monte S.Angelo, 80126 Napoli, Italy

[91] NASA Goddard Space Flight Center, Greenbelt, MD 20771, USA

[92] Dipartimento di Fisica, Università degli Studi di Genova, 16146 Genoa, Italy

[93] Tsinghua University, Beijing 100084, China

[94] Texas Tech University, Lubbock, TX 79409, USA

[95] The University of Mississippi, University, MS 38677, USA

[96] Museo Storico della Fisica e Centro Studi e Ricerche “Enrico Fermi”, 00184 Rome, Italy

[97] The Pennsylvania State University, University Park, PA 16802, USA

[98] Department of Applied Physics, Fukuoka University, Nanakuma, Jonan, Fukuoka 814-0180, Japan

[99] National Tsing Hua University, Hsinchu City 30013, Taiwan, ROC

[100] Charles Sturt University, Wagga Wagga, New South Wales 2678, Australia

[101] Canadian Institute for Theoretical Astrophysics, University of Toronto, Toronto, Ontario M5S 3H8, Canada

[102] Tamkang University, No.151, Yingzhuan Rd., Danshui Dist., New Taipei City 25137, Taiwan

[103] University of Chicago, Chicago, IL 60637, USA

[104] Department of Physics, The Center for High Energy and High Field Physics, National Central University, No. 300, Zhongda Road, Zhongyi District, Taiyuan City 320, Taiwan

[105] Department of Physics, National Tsing Hua University, No. 101, Section 2, Kuang-Fu Road, Hsinchu 30013, Taiwan

[106] Institute of Astronomy, National Tsing Hua University, No. 101, Section 2, Kuang-Fu Road, Hsinchu 30013, Taiwan

[107] The Chinese University of Hong Kong, Shatin, NT, Hong Kong

[108] Seoul National University, Seoul 08826, South Korea

[109] Pusan National University, Busan 46241, South Korea

[110] Carleton College, Northfield, MN 55057, USA

[111] Department of Physics, National Taiwan Normal University, 88 section 4 Ting-Chou Rd., Taipei 116, Taiwan

[112] INAF, Osservatorio Astronomico di Padova, 35122 Padova, Italy

[113] INFN, Trento Institute for Fundamental Physics and Applications, 38123 Povo, Trento, Italy

[114] OzGrav, University of Melbourne, Parkville, VIC 3010, Australia

[115] Columbia University, New York, NY 10027, USA

[116] Universitat de les Illes Balears IAC3—IEEC, 07122 Palma de Mallorca, Spain

[117] Université Libre de Bruxelles, 1050 Brussels, Belgium

[118] Sonoma State University, Rohnert Park, CA 94928, USA

[119] Departamento de Matemáticas, Universitat de València, 46100 Burjassot, València, Spain

[120] University of Rhode Island, Kingston, RI 02881, USA

[121] The University of Texas Rio Grande Valley, Brownsville, TX 78520, USA

[122] Bellevue College, Bellevue, WA 98007, USA

[123] MTA-ELTE Astrophysics Research Group, Institute of Physics, Eötvös University, Budapest 1117, Hungary

[124] Institute for Plasma Research, Bhat, Gandhinagar 382428, India

[125] The University of Sheffield, Sheffield S10 2TN, UK

[126] IGFAE, Campus Sur, Universidade de Santiago de Compostela, 15782 Santiago de Compostela, Spain

[127] Dipartimento di Scienze Matematiche, Fisiche e Informatiche,Università di Parma, 43124 Parma, Italy

[128] California State University, Los Angeles, 5151 State University Dr., Los Angeles, CA 90032, USA

[129] Dipartimento di Fisica, Università di Trento, 38123 Povo, Trento, Italy

[130] Università di Roma 'La Sapienza', 00185 Roma, Italy

[131] Department of Physics, University of Toyama, 3190 Gofuku, Toyama-shi, Toyama 930-8555, Japan

[132] Colorado State University, Fort Collins, CO 80523, USA

[133] Kenyon College, Gambier, OH 43022, USA

[134] Christopher Newport University, Newport News, VA 23606, USA

[135] Observatori Astronòmic, Universitat de València, 46980 Paterna, València, Spain

[136] Department of Astronomy, The University of Tokyo, 2-21-1, Osawa, Mitaka-shi, Tokyo 181-8588, Japan

[137] Advanced Technology Center, National Astronomical Observatory of Japan (NAOJ), 2-21-1, Osawa, Mitaka-shi, Tokyo 181-8588, Japan

[138] School of Mathematics, University of Edinburgh, Edinburgh EH9 3FD, UK

[139] Institute Of Advanced Research, Gandhinagar 382426, India

[140] Indian Institute of Technology Bombay, Powai, Mumbai 400076, India

[141] Wuhan Institute of Physics and Mathematics, Chinese Academy of Sciences, West No. 30, Xiaohongshan, Wuhan 430071, China

[142] University of Szeged, Dóm tér 9, Szeged 6720, Hungary

[143] Tata Institute of Fundamental Research, Mumbai 400005, India

[144] INAF, Osservatorio Astronomico di Capodimonte, 80131 Napoli, Italy

[145] University of Michigan, Ann Arbor, MI 48109, USA

[146] Applied Research Laboratory, High Energy Accelerator Research Organization (KEK), 1-1, Oho, Tsukuba-shi, Ibaraki 305-0801, Japan

[147] Academia Sinica, Institute of Physics, 128 Sec. 2, Academia Rd., Nankang, Taipei 11529, Taiwan

[148] American University, Washington, DC 20016, USA

[149] Institute for Cosmic Ray Research (ICRR), The University of Tokyo, Higashi-Mozumi 238, Kamioka-cho, Hida-shi, Gifu 506-1205, Japan

[150] College of Industrial Technology, Nihon University, 1-2-1,Izumi-cho, Narashino-shi, Chiba 275-8575, Japan

[151] GRAPPA, Anton Pannekoek Institute for Astronomy and Institute of High-Energy Physics, University of Amsterdam, Science Park 904, 1098 XH Amsterdam, The Netherlands

[152] Delta Institute for Theoretical Physics, Science Park 904, 1090 GL Amsterdam, The Netherlands

[153] Graduate School of Science and Technology, Niigata University, 8050, Ikarashi-2-no-cho, Nishi-ku, Niigata-shi, Niigata 950-2181, Japan

[154] Institute for Cosmic Ray Research (ICRR), Research Center for Cosmic Neutrinos (RCCN), The University of Tokyo, 5-1-5 Kashiwa-no-Ha, Kashiwa City, Chiba 277-8582, Japan

[155] Kavli Institute for Astronomy and Astrophysics, Peking University, Yiheyuan Road 5, Haidian District, Beijing 100871, China

[156] Yukawa Institute for Theoretical Physics (YITP), Kyoto University, Oiwake-cho, KitaShirakawa, Sakyou-ku, Kyoto-shi, Kyoto 606-8502, Japan

[157] Graduate School of Science, Osaka City University, 3-3-138, Sugimoto-cho, Sumiyosi-ku, Osaka-shi, Osaka 558-8585, Japan

[158] Nambu Yoichiro Institute of Theoretical and Experimental Physics (NITEP), Osaka City University, 3-3-138, Sugimoto-cho,Sumiyosi-ku, Osaka-shi, Osaka 558-8585, Japan

[159] JAXA Institute of Space and Astronautical Science, 3-1-1 Yoshinodai, Chuo-ku, Sagamihara, Kanagawa, Japan

[160] Directorate of Construction, Services and Estate Management, Mumbai 400094, India

[161] University of Białystok, 15-424 Białystok, Poland

[162] King's College London, University of London, London WC2R 2LS, UK

[163] University of Southampton, Southampton SO17 1BJ, UK

[164] Department of Physics, Ulsan National Institute of Science and Technology (UNIST), 50 UNIST-gil, Ulju-gun, Ulsan 44919, South Korea

[165] Graduate School of Science and Technology, Tokyo Institute of Technology, 2-12-1, Ookayama, Meguro-ku, Tokyo 152-8551, Japan

[166] University of Washington Bothell, Bothell, WA 98011, USA

[167] Institute of Applied Physics, Nizhny Novgorod, Russia 603950

[168] Ewha Womans University, Seoul 03760, South Korea

[169] Department of Physics, Ewha Womans University, 52, Ewhayeodae-gil, Seodaemun-gu, Seoul 03760, Korea

[170] Inje University, Gimhae, South Gyeongsang 50834, South Korea

[171] Department of Computer Simulation, Inje University, 197 Inje-ro, Gimhae, Gyeongsangnam-do 50834, Korea

[172] Universität Hamburg, 22761 Hamburg, Germany

[173] Maastricht University, P.O. Box 616, 6200 MD Maastricht, The Netherlands

[174] Department of Physical Science, Hiroshima University, 1-3-1, Kagamiyama, Higashihiroshima-shi, Hiroshima 903-0213, Japan

[175] Institute for Cosmic Ray Research (ICRR), Research Center for Cosmic Neutrinos (RCCN), The University of Tokyo, Higashi-Mozumi 238, Kamioka-cho, Hida-shi, Gifu 506-1205, Japan

[176] NCBJ, 05-400 Świerk-Otwock, Poland

[177] Institute of Mathematics, Polish Academy of Sciences, 00656 Warsaw, Poland

[178] Cornell University, Ithaca, NY 14850, USA

[179] Institute for Advanced Research, Nagoya University, Furocho, Chikusa-ku, Nagoya, Aichi 464-8602, Japan

[180] Hillsdale College, Hillsdale, MI 49242, USA

[181] Hanyang University, Seoul 04763, South Korea

[182] Korea Astronomy and Space Science Institute, Daejeon 34055, South Korea

[183] National Applied Research Laboratories, National Center for High-performance computing, No. 7, R&D 6th Rd., Hsinchu Science Park, Hsinchu City 30076, Taiwan

[184] NASA Marshall Space Flight Center, Huntsville, AL 35811, USA

[185] Dipartimento di Matematica e Fisica, Università degli Studi Roma Tre, 00146 Rome, Italy

[186] INFN, Sezione di Roma Tre, 00146 Rome, Italy

[187] ESPCI, CNRS, 75005 Paris, France

[188] OzGrav, Swinburne University of Technology, Hawthorn, VIC 3122, Australia

[189] University of Portsmouth, Portsmouth PO1 3FX, UK

[190] Southern University and A&M College, Baton Rouge, LA 70813, USA

[191] College of William and Mary, Williamsburg, VA 23187, USA

[192] Centre Scientifique de Monaco, 8 quai Antoine Ier, 98000 Monaco, Monaco

[193] Institute for Photon Science and Technology, The University of Tokyo, 2-11-16, Yayoi, Bunkyo-ku, Tokyo 113-8656, Japan

[194] Indian Institute of Technology Madras, Chennai 600036, India

[195] The Applied Electromagnetic Research Institute, National Institute of Information and Communications Technology (NICT), 4-2-1, Nukuikita-machi, Koganei-shi, Tokyo 184-8795, Japan

[196] INFN Sezione di Torino, Via P. Giuria 1, 10125 Torino, Italy

[197] Institut des Hautes Etudes Scientifiques, 91440 Bures-sur-Yvette, France

[198] Faculty of Law, Ryukoku University, 67 Fukakusa Tsukamoto-cho, Fushimi-ku, Kyoto 612-8577, Japan

[199] Department of Physics, Kyoto University, Oiwake-cho, KitaShirakawa, Sakyou-ku, Kyoto-shi, Kyoto 606-8502, Japan

[200] IISER-Kolkata, Mohanpur, West Bengal 741252, India

[201] Kobayashi-Maskawa Institute for the Origin of Particles and the Universe, Nagoya University, Furocho, Chikusa-ku, Nagoya, Aichi 464-8602, Japan

[202] Whitman College, 345 Boyer Avenue, Walla Walla, WA 99362, USA

[203] Faculty of Engineering, Niigata University, 8050, Ikarashi-2-no-cho, Nishi-ku, Niigata-shi, Niigata 950-2181, Japan

[204] Department of Physics, Sogang University, One Sinsu-Dong, Mapo-Gu, Seoul 121-742, Korea

[205] Université de Lyon, 69361 Lyon, France

[206] Hobart and William Smith Colleges, Geneva, NY 14456, USA

[207] Department of Engineering, University of Sannio, 82100 Benevento, Italy

[208] Janusz Gil Institute of Astronomy, University of Zielona Góra, 65-265 Zielona Góra, Poland

[209] University of Washington, Seattle, WA 98195, USA

[210] SUPA, University of the West of Scotland, Paisley PA1 2BE, UK

[211] Faculty of Arts and Science, Kyushu University, 744, Motooka, Nishi-ku, Fukuoka 819-0395, Japan

[212] Department of Electronic Control Engineering, Nagaoka College, National Institute of Technology, 888 Nishikatakai, Nagaoka, Niigata 940-8532, Japan

[213] Graduate School of Science and Engineering, Hosei University, 3-7-2, Kajino-cho, Koganei-shi, Tokyo 184-8584, Japan

[214] Faculty of Science, Toho University, 2-2-1 Miyama, Funabashi-shi, Chiba, Japan

[215] Indian Institute of Technology Gandhinagar, Ahmedabad, Gujarat 382424, India

[216] Université de Montréal/Polytechnique, Montreal, QC H3T 1J4, Canada

[217] Faculty of Information Science and Technology, Osaka Institute of Technology, Kitayama 1-79-1, Hirakata, Osaka 573-0196, Japan

[218] Indian Institute of Technology Hyderabad, Sangareddy, Khandi, Telangana 502285, India

[219] Division of Theoretical Astronomy, National Astronomical Observatory of Japan (NAOJ), 2-21-1, Osawa, Mitaka-shi, Tokyo 181-8588, Japan

[220] International Institute of Physics, Universidade Federal do Rio Grande do Norte, Natal, RN 59078-970, Brazil

[221] Villanova University, 800 Lancaster Ave, Villanova, PA 19085, USA

[222] Andrews University, Berrien Springs, MI 49104, USA

[223] Department of Information and Management Systems Engineering, Nagaoka University of Technology, 1603-1 Kamitomioka, Nagaoka, Niigata 940-2188, Japan

[224] National Institute of Advanced Industrial Science and Technology, National Metrology Institute of Japan, 1-1-1, Umezono, Tsukuba-shi, Ibaraki 305-8568, Japan

[225] Max Planck Institute for Gravitational physik (Albert Einstein Institute), 14476 Potsdam-Golm, Germany

[226] Università di Siena, 53100 Siena, Italy

[227] Faculty of Science, Niigata University, 8050, Ikarashi-2-no-cho, Nishi-ku, Niigata-shi, Niigata 950-2181, Japan

[228] Department of Communications, National Defense Academy of Japan, Hashirimizu 1-10-20, Yokosuka-shi, Kanagawa-Pref 239-8686, Japan

[229] Trinity University, San Antonio, TX 78212, USA

[230] Van Swinderen Institute for Particle Physics and Gravity, University of Groningen, Nijenborgh 4, 9747 AG Groningen, The Netherlands

[231] Department of Physics and Astronomy, Sejong University, 209 Neungdong-ro, Gwangjin-gu, Seoul 143-747, South Korea

[232] Department of Astronomy, Beijing Normal University, Beijing 100875, China

[233] Department of Physics, University of Texas, Austin, TX 78712, USA